\documentstyle[aps,12pt,preprint,psfig]{revtex}
\tightenlines

\begin{document}
\preprint{
\hbox{PKU-TP-98-53}}
\draft

\title{Diffractive light quark jet and gluon jet production at
        hadron colliders in the two-gluon exchange model}

\author{Feng Yuan}
\address{\small {\it Department of Physics, Peking University, Beijing 100871, People's Republic
of China}}
\author{Kuang-Ta Chao}
\address{\small {\it China Center of Advanced Science and Technology (World Laboratory), Beijing 100080,
        People's Republic of China\\
      and Department of Physics, Peking University, Beijing 100871, People's Republic of China}}

\maketitle
\begin{abstract}

Massless diquark jet and gluon jet productions at large transverse momentum in
the coherent diffractive processes at hadron colliders are studied in the
two-gluon exchange parametrization of the Pomeron model. We use the helicity
amplitude method to calculate the cross section formulas. We find that for the
light quark jet and gluon jet production the diffractive process is related to the
differential off-diagonal gluon distribution function in the proton.
We estimate the production rate for these processes at the Fermilab Tevatron
by approximating the off-diagonal gluon distribution function by the usual
diagonal gluon distribution in the proton.
We also use the helicity amplitude method to calculate the diffractive
charm jet production at hadron colliders, by which we reproduce the leading
logarithmic approximation result of this process we previously calculated.
\end{abstract}

\pacs{PACS number(s): 12.40.Nn, 13.85.Ni, 14.40.Gx}

\section{Introduction}

In recent years, there has been a renaissance of interest in
diffractive scattering.
These diffractive processes are described by the Regge theory in
terms of the Pomeron ($I\!\! P$) exchange\cite{pomeron}.
The Pomeron carries quantum numbers of the vacuum, so it is a colorless entity
in QCD language, which may lead to the ``rapidity gap" events in experiments.
However, the nature of Pomeron and its reaction with hadrons remain a mystery.
For a long time it had been understood that the dynamics of the
``soft pomeron'' is deeply tied to confinement.
However, it has been realized now that how much can be learned about
QCD from the wide variety of small-$x$ and hard diffractive processes,
which are now under study experimentally.
In Refs.\cite{th1,th2}, the diffractive $J/\psi$ and $\Upsilon$ production
cross section have been formulated in photoproduction processes and in
DIS processes in perturbative QCD.
In the framework of perturbative QCD the Pomeron is represented by a pair of
gluon in the color-singlet sate.
This two-gluon exchange model can successfully describe the experimental
results from HERA\cite{hera-ex}.

On the other hand, as we know that there exist nonfactorization effects
in the hard diffractive processes at hadron colliders
\cite{preqcd,collins,soper,tev}.
First, there is the so-called spectator effect\cite{soper}, which can
change the probability of the diffractive hadron emerging from collisions
intact. Practically, a suppression factor (or survive factor) ``$S_F$"
is used to describe this effect\cite{survive}.
Obviously, this suppression factor can not be calculated in perturbative
QCD, which is now viewed as a nonperturbative parameter.
Typically, the suppression factor $S_F$ is determined to be about
$0.1$ at the energy scale of the Fermilab Tevatron\cite{tev}.
Another nonfactorization effect discussed in literature is associated with the coherent
diffractive processes at hadron colliders\cite{collins}, in which
the whole Pomeron is induced in the hard scattering.
It is proved in \cite{collins} that the existence of the leading twist
coherent diffractive processes is associated with a breakdown of the
QCD factorization theorem.

Based on the success of the two-gluon exchange parametrization of the Pomeron
model in the description of the diffractive photoproduction processes
at $ep$ colliders\cite{th1,th2,hera-ex}, we may extend the applications
of this model to calculate the diffractive processes at hadron colliders in perturbative QCD.
Under this context, the Pomeron represented by a color-singlet two-gluon
system emits from one hadron and interacts with another hadron in hard
process, in which the two gluons are both involved (see Fig.~1).
Therefore, these processes calculated in the two-gluon exchange model are just
belong to the coherent diffractive processes in hadron collisions.
Another important feature of the calculations of the diffractive processes
in this model recently demonstrated is the sensitivity
to the off-diagonal parton distribution
function in the proton\cite{offd}.

Using this two-gluon exchange model, we have calculated the  diffractive
$J/\psi$ production \cite{psi}, charm jet production\cite{charm},
massive muon pair and $W$ boson productions\cite{dy} in hadron collisions.
These calculations show that we can explore the off-diagonal
gluon distribution function in low $x$ region and
study the coherent diffractive processes
at hadron colliders through these processes.
In this paper, we will calculate the light quark (massless) jet and gluon jet
productions at large transverse momentum in the coherent diffractive processes
 at hadron colliders by using the two-gluon exchange model.
As sketched in Fig.~1, the two-gluon system (in color-singlet) emitted from one hadron
interacts with another hadron to produce the final state light
quark (or gluon) jets.
There are three different partonic processes contributing to the diffractive dijet productions,
$gp\rightarrow q\bar qp$, $qp\rightarrow qgp$, and $gp\rightarrow ggp$
as shown in Fig.2, Fig.3 and Fig.4 respectively.

In the calculations of Refs.\cite{psi,charm,dy}, there always is a large mass
scale associated with the production process.
That is $M_\psi$ for $J/\psi$ production, $m_c$ for the charm jet
production, $M^2$ for the massive muon production ($M^2$ is the invariant mass
of the muon pair) and $M_W^2$ for $W$ boson production.
However, in the processes we will calculate in this paper, 
there is no large mass scale.
So, for the light quark jet and gluon jet production, the large transverse momentum is needed to
guarantee the application of the perturbative QCD.
Furthermore, the experience on the calculations of the diffractive di-quark jet
photoproduction\cite{zaka} shows that the light quark jet production
in the two-gluon exchange model has a distinctive feature that
there is no contribution from the small $l_T^2$ region ($l_T^2<k_T^2$)
in the integration of the amplitude over $l_T^2$.
So, the expansion (in terms of $l_T^2/M_X^2$) method 
used in Refs.\cite{psi,charm,dy} can not be applied for the calculations
here.
In the following calculations, we will employ the helicity amplitude method
to calculate the amplitude of the diffractive light quark jet and gluon jet
production in hadron collisions.
We will show that the production cross sections for these processes
are related to the differential (off-diagonal) gluon distribution
function in the proton as that in the diffractive di-quark jet
photoprodution process\cite{zaka}.
(In contrast, we note that the cross sections of the processes
calculated in Refs.\cite{psi,charm,dy} are related to 
the integrated gluon distribution function in the proton).

The diffractive production of heavy quark jet at hadron colliders has also
been studied by using the two-gluon exchange model in Ref.\cite{levin}.
However, their approach is very different from ours
\footnote{For detailed discussions and comments, please see \cite{charm}}.
In their calculations, they separated their diagrams into two parts,
and called one part the coherent diffractive contribution to the heavy
quark production.
In our approach, in line with the definition of
Ref.\cite{collins}, we call
the process in which the whole Pomeron participates in the hard scattering
process as the coherent diffractive process.
Under this definition, all of the diagrams plotted in Fig.2, Fig.3 and Fig.4
for the partonic processes $gp\rightarrow q\bar qp$, $qp\rightarrow qgp$
and $gp\rightarrow ggp$ contribute to
the coherent diffractive production.

In addition, we must emphasize that in our calculations we only discuss the
contribution to the
diffractive dijet production from the so-called ``Pomeron fragmentation"
region, i.e., $M_X^2\sim 4k_T^2$, where $M_X^2$ is the invariant mass of the
diffractive final states and $k_T$ is the transverse momenta of the outgoing
jet. However, in the region of $M_X^2\gg 4k_T^2$, the diffractive
production may be dominated by the central (for $M^2_X$ system) or initial
hadron fragmentation contributions, and in this region a resolved Pomeron model
(with parton distributions for the Pomeron) may be more appropriate.

The rest of the paper is organized as follows.
In Sec.II, we will give the cross section formulas for the partonic processes
of Figs.2-4 in the leading order of perturbative QCD,
where we employ the helicity amplitude method to calculate the amplitude
for these processes.
We will also use this method to recalculate the diffractive
charm jet production process $gp\rightarrow c\bar c p$, by which
we can reproduce the leading logarithmic approximation result given in Ref.\cite{charm}.
In Sec.III, the numerical results for the diffractive light quark jet and
gluon jet production at the Fermilab Tevatron will be given.
And the conclusions will be given in Sec.IV.

\section{ The cross section formulas for the partonic processes}

In the leading order of perturbative QCD,
the Feynman diagrams for the three different partonic
processes are plotted in Figs.~2-4 respectively.
The two-gluon system coupled to the proton (antiproton) in these diagrams
 is in a color-singlet state, which characterizes the diffractive processes in
perturbative QCD.
Due to the positive signature of these diagrams (color-singlet exchange),
we know that the real part of the amplitude cancels out in the leading
logarithmic approximation.
To get the imaginary part of the amplitude, we must calculate the
discontinuity represented by the crosses in each diagram of these figures.

In our calculations, we express the formulas in terms of the Sudakov
variables.
That is, every four-momenta $k_i$ are decomposed as,
\begin{equation}
k_i=\alpha_i q+\beta_i p+\vec{k}_{iT},
\end{equation}
where $q$ and $p$ are the momenta of the incident parton (quark or gluon) and
the diffractive proton,
$q^2=0$, $p^2=0$, and $2p\cdot q=W^2=s$.
$\alpha_i$ and $\beta_i$ are the momentum fractions of $q$ and $p$
respectively.
$k_{iT}$ is the transverse momentum, which satisfies
\begin{equation}
k_{iT}\cdot q=0,~~~
k_{iT}\cdot p=0.
\end{equation}

All of the Sudakov variables for every momentum
are determined by using the on-shell conditions
of the momenta represented by the external lines and the crossed lines in the diagram.
The calculations of these Sudakov variables are similar to those in
the diffractive charm jet production process $gp\rightarrow c\bar c p$\cite{charm}.

For the momentum $u$, we have
\begin{equation}
\alpha_u=0,~~\beta_u=x_{I\! P}=\frac{M_X^2}{s},~~u_T^2=t=0,
\end{equation}
where $M_X^2$ is the invariant mass squared of the diffractive final state.
For the high energy diffractive process, we know that $M_X^2\ll s$, so
we have $\beta _u$ ($x_{I\! P}$) as a small parameter.
For the momentum $k$,
\begin{equation}
\label{ak}
\alpha_k(1+\alpha_k)=-\frac{k_T^2}{M_X^2},~~\beta_k=-\alpha_k\beta_u,
\end{equation}
where $k_T$ is the transverse momentum of the out going quark jet (or gluon jet).

For the loop mentum $l$, the Sudakov parameters are not the same for every diagrams
and every partonic processes,
so we will discuss them in the following subsections respectively.

By the form of these Sudakov variables, the cross section formula for the
partonic process can be formulated as
\begin{equation}
\label{xs}
\frac{d\hat{\sigma}}{dt}|_{t=0}=\frac{dM_X^2d^2k_Td\alpha_k}{16\pi s^216\pi^3M_X^2}
        \delta(\alpha_k(1+\alpha_k)+\frac{k_T^2}{M_X^2})\sum \overline{|{\cal A}|}^2.
\end{equation}
${\cal A}$ is the amplitude of the partonic process, and we know that the real part
of the amplitude is zero in the leading order.
So, we only need to calculate the imaginary parts of the amplitudes for the
three partonic processes.

In the calculations of the amplitudes for the three partonic processes,
we employ the {\it Helicity Amplitude} method\cite{wu,ham1,ham}.
As a cross check, we will recalculate the diffractive charm jet production\cite{charm}
by using this method.

Another important check to our calculations is performed by examining the
behavior of the amplitude under the limit of $l_T^2\rightarrow 0$ in the
integration over the loop momentum $l$.
From the experience of the calculations for the diffractive charm jet
production in Ref.\cite{charm}, we know that the contribution to the amplitude
from the individual diagram of the parton process has linear singularity
at $l_T^2\rightarrow 0$. The linear singularity is not proper
in QCD perturbative calculations. So, the linear singularities coming from different
diagrams must be canceled out by each other for the same order of perturbative
calculations.
That is to say we must guarantee the total sum of the contributions from
all of the diagrams free of the linear singularity in the calculations
for every partonic processes.

\subsection{$gp\rightarrow q\bar qp$ process}

The first four diagrams of Fig.~2 for this subprocess are the same as those
calculated in the diffractive photoproduction processes\cite{th1,th2,th3}.
But, due to the existence of
gluon-gluon interaction vertex in QCD, in the partonic process
$gp\rightarrow q\bar q p$, there are additional five diagrams (Diags.~(5)-(9)).
These five diagrams are needed for complete calculations
in this order of QCD.

First, we give the Sudakov parameters for the loop mentum $l$ for this
process (Fig.~2),
\begin{eqnarray}
\nonumber
\alpha_l&=&-\frac{l_T^2}{s},\\
\nonumber
\beta_l&=&\frac{2(k_T,l_T)-l_T^2}{\alpha_ks},~~~{\rm for~Diag.}1,~3,~5,\\
\nonumber
&=&\frac{2(k_T,l_T)+l_T^2}{(1+\alpha_k)s},~~~{\rm for~Diag.}2,~4,~6,\\
&=&-\frac{M_X^2-l_T^2}{s},~~~~~~~{\rm for~Diag.}7,~8,~9,
\end{eqnarray}
where $(k_T,l_T)$ is the 2-dimensional product of the transverse vectors
$\vec{k}_T$ and $\vec{l}_T$.

The imaginary part of the amplitude ${\cal A}(gp\rightarrow q\bar qp)$ for each diagram
of Fig.~2 has the following general form,
\begin{equation}
\label{ima}
{\rm Im}{\cal A}=C_F(T_{ij}^a)\int \frac{d^2l_T}{(l_T^2)^2}F\times\bar u
        _i(k+q)\Gamma_\mu v_j(u-k),
\end{equation}
where $C_F$ is the color factor for each diagram, and is the same as that
for the $gp\rightarrow c\bar c p$ process\cite{charm}.
$a$ is the color index of the incident gluon.
$\Gamma_\mu$ represents some $\gamma$ matrices including one
propagator. $F$ in the integral is defined as
\begin{equation}
\label{feq}
F=\frac{3}{2s}g_s^3f(x',x^{\prime\prime};l_T^2),
\end{equation}
where
\begin{equation}
\label{offd1}
f(x',x^{\prime\prime};l_T^2)=\frac{\partial G(x',x^{\prime\prime};l_T^2)}{\partial {\rm ln} l_T^2},
\end{equation}
where the function
$G(x',x^{\prime\prime};k_T^2)$ is the so-called
off-diagonal gluon distribution function\cite{offd}.
Here, $x'$ and $x^{\prime\prime}$ are the momentum fractions of the proton
carried by the two gluons.
It is expected that at small $x$, there is no big difference between the off-diagonal and
the usual diagonal gluon densities\cite{off-diag}.
So, in the following calculations, we estimate the production rate by
approximating the off-diagonal gluon density by 
the usual diagonal gluon density, 
$G(x',x^{\prime\prime};Q^2)\approx xg(x,Q^2)$, where $x=x_{I\!\! P}=M_X^2/s$.

In Ref.\cite{charm}, we calculate the amplitude (\ref{ima}) for the heavy quark
diffractive production processes by expanding
$\Gamma_\mu$ in terms of $l_T^2$. However, in the light quark jet production
process $gp\rightarrow q\bar qp$, the expansion method is not yet valid.
According to the result of \cite{charm}, the production cross section
is proportional to the heavy quark mass. If we apply this formula to the
light quark jet production, the cross section will be zero.
That is to say, the expansion of the amplitude in terms of $l_T^2$, in which
the large logarithmic contribution comes from the region of $l_T^2\ll M_X^2$,
is not further suitable for the calculation of the cross section for massless light
quark jet production.
Furthermore, the experience of the calculation of the diffractive light quark
jet photoproduction process $\gamma p\rightarrow q\bar qp$ (for $Q^2=0$)
\cite{zaka} indicates that there is no contribution from
the region of $l_T^2< k_T^2$ in the integration of the amplitude over
$l_T^2$.
In the hadroproduction process $gp\rightarrow q\bar qp$,
the situation is the same.
So, the expansion method, in which $l_T^2$ is taken as a small
parameter, is not valid for the calculations of the massless quark production
processes.

In the following, we employ the helicity amplitude method\cite{ham} to calculate
the amplitude Eq.(\ref{ima}). For the massless quark spinors, we define
\begin{equation}
\label{spinor}
u_\pm(p)=\frac{1}{\sqrt{2}}(1\pm\gamma_5)u(p).
\end{equation}
For the polarization vector of the incident gluon, which is transversely polarized,
we choose,
\begin{equation}
\label{ev}
e_\pm=\frac{1}{\sqrt{2}}(0,1,\pm i,0).
\end{equation}
The helicity amplitudes for the processes in which the polarized Dirac particles are involved
have the following general forms\cite{ham},
\begin{equation}
\label{ham}
\bar u_\pm(p_f)Qv_\mp(p_i)=\frac{Tr[Q\not\! p_i\not\! n\not\! p_f(1\mp\gamma_5)]}
        {4\sqrt{(n\cdot p_i)(n\cdot p_f)}},
\end{equation}
where $n$ is an arbitrary massless 4-vector, which is set to be $n=p$ in the
following calculations.
Using this formula (\ref{ham}), the calculations of the helicity amplitude
${\cal A}(\lambda(g),\lambda(q),\lambda({\bar q}))$ for the diffractive
process $gp\rightarrow q\bar q p$ is straightforward.
Here $\lambda$ are the corresponding helicities of the external gluon,
quark and antiquark.
In our calculations, we only take the leading order contribution, and neglect the
higher order contributions which are proportional to $\beta_u=\frac{M_X^2}{s}$
because in the high energy diffractive processes we have $\beta_u\ll 1$.

For the first four diagrams, to sum up together, the imaginary part of the
amplitude ${\cal A}(\pm,+,-)$ is
\begin{equation}
\label{im1}
{\rm Im}{\cal A}^{1234}(\pm,+,-)=\alpha_k^2(1+\alpha_k){\cal N}\times
        \int\frac{d^2\vec{l}_T}{(l_T^2)^2}f(x',x'';l_T^2)
        (-\frac{2}{9}\frac{\vec{e}^{(\pm)}\cdot \vec{k}_T}{k_T^2}-\frac{1}{36}\frac{\vec{e}^{(\pm)}\cdot(\vec{k}_T-\vec{l}_T)}{(\vec{k}_T-\vec{l}_T)^2}),
\end{equation}
where $\frac{2}{9}$ and $-\frac{1}{36}$ are the color factors for Diag.1,4 and
Diag.2,3 respectively, and
${\cal N}$ is defined as
\begin{equation}
\label{ne}
{\cal N}=\frac{s}{\sqrt{-\alpha_k(1+\alpha_k)}}g_s^3T_{ij}^a.
\end{equation}
The other helicity amplitudes for the first four diagrams have the similar form
as (\ref{im1}).
The amplitude expression Eq.~(\ref{im1}) is the same as that for the photoproduction
process $\gamma p\rightarrow q\bar q p$ previously calculated in Refs.\cite{th3,zaka} except the difference
on the color factors. In the diffractive photoproduction process, the color
factors of these four diagrams are the same (they are all $\frac{2}{9}$), while in hadroproduction
process the color factors are no longer the same for these four diagrams.
It is instructive to see what is the consequence of this difference.
We know that the amplitude of the diffractive process in Eq.~(\ref{ima}) must be
zero in the limit $l_T^2\rightarrow 0$. Otherwise, this will lead to a linear singularity
when we perform the integration of the amplitude over $l_T^2$ due to existence of the factor $1/(l_T^2)^2$ in the
integral of Eq.~(\ref{ima})\cite{charm}.
This linear singularity is not proper in QCD calculations.
So, we must first exam the amplitude behavior under the limit of $l_T^2\rightarrow 0$
for all the diffractive processes in the calculations using the two-gluon exchange model.
From Eq.~(\ref{im1}), we can see that the amplitude for the
diffractive photoproduction of di-quark jet process is exact zero at $l_T^2\rightarrow 0$.
However, for the diffractive hadroproduction process $gp\rightarrow q\bar qp$ the amplitude
for the first four diagrams 
is not exact zero in the limit $l_T^2\rightarrow 0$ due to the inequality of the
color factors between them.
So, for $gp\rightarrow q\bar qp$ process there must be other diagrams in this order
of perturbative QCD calculation to cancel out the linear singularity which
rises from the first four diagrams.
The last five diagrams of Fig.2 are just for this purpose.

For example, the contributions from Diags.5 and 8 are
\begin{equation}
{\rm Im}{\cal A}^{58}(\pm,+,-)=\alpha_k^2(1+\alpha_k){\cal N}\times
        \int\frac{d^2\vec{l}_T}{(l_T^2)^2}f(x',x'';l_T^2)
        (-\frac{1+\alpha_k}{4}\frac{\vec{e}^{(\pm)}\cdot (\vec{k}_T-(1+\alpha_k)\vec{l}_T)}{(\vec{k}_T-(1+\alpha_k)^2\vec{l}_T)^2}),
\end{equation}
and the contributions from Diags.6 and 9 are 
\begin{equation}
{\rm Im}{\cal A}^{69}(\pm,+,-)=\alpha_k^2(1+\alpha_k){\cal N}\times
        \int\frac{d^2\vec{l}_T}{(l_T^2)^2}f(x',x'';l_T^2)
        (\frac{\alpha_k}{4}\frac{\vec{e}^{(\pm)}\cdot (\vec{k}_T-\alpha_k\vec{l}_T)}{(\vec{k}_T-\alpha_k^2\vec{l}_T)^2}),
\end{equation}
and the contribution from Diag.7 is
\begin{equation}
{\rm Im}{\cal A}^{7}(\pm,+,-)=\alpha_k^2(1+\alpha_k){\cal N}\times
        \int\frac{d^2\vec{l}_T}{(l_T^2)^2}f(x',x'';l_T^2)
        (\frac{1}{2}\frac{\vec{e}^{(\pm)}\cdot \vec{k}_T}{k_T^2}).
\end{equation}
From the above results, we can see that the contributions from Diags.5-9
just cancel out the linear singularity which rises from the first four diagrams.
Their total sum from the nine diagrams of Fig.2 is free of linear
singularity now.

Finally, by adding up all of the nine diagrams of Fig.2, the
imaginary parts of the amplitudes for the following helicity sets
are,
\begin{eqnarray}
\nonumber
{\rm Im}{\cal A}(\pm,+,-)&=&\alpha_k^2(1+\alpha_k){\cal N}\times {\cal T}^{(\pm)},\\
{\rm Im}{\cal A}(\pm,-,+)&=&\alpha_k(1+\alpha_k)^2{\cal N}\times {\cal T}^{(\pm)},
\end{eqnarray}
where
\begin{eqnarray}
\label{int2}
\nonumber
{\cal T}^{(\pm)}&=&\int\frac{d^2\vec{l}_T}{(l_T^2)^2}f(x',x'';l_T^2)[(\frac{1}{2}-\frac{2}{9})
        \frac{\vec{e}^{(\pm)}\cdot \vec{k}_T}{k_T^2}-\frac{1}{36}\frac{\vec{e}^{(\pm)}\cdot(\vec{k}_T-\vec{l}_T)}{(\vec{k}_T-\vec{l}_T)^2}\\
&&-\frac{1+\alpha_k}{4}\frac{\vec{e}^{(\pm)}\cdot(\vec{k}_T-(1+\alpha_k)\vec{l}_T)}{(\vec{k}_T-(1+\alpha_k)\vec{l}_T)^2}
  +\frac{\alpha_k}{4}\frac{\vec{e}^{(\pm)}\cdot(\vec{k}_T-\alpha_k\vec{l}_T)}{(\vec{k}_T-\alpha_k\vec{l}_T)^2}].
\end{eqnarray}
And the amplitudes for the other helicity sets are zero in the light quark jet
production process.
From the above results, we can see that in the integration of the amplitude
the linear singularity from different diagrams are canceled out by each other,
which will guarantee there is no linear singularity in the total sum.

Another feature of the above results for the amplitudes is the relation to the
differential off-diagonal gluon distribution function $f(x',x'';l_T^2)$.
To get the cross section for the diffractive process, we must perform the
integration of Eq.~(\ref{int2}).
However, as mentioned above that there is no big difference between the off-diagonal
gluon distribution function and the usual gluon distribution at small $x$,
so we can simplify the integration of (\ref{int2}) by 
approximating the differential off-diagonal gluon
distribution function $f(x',x'';l_T^2)$ by the usual diagonal differential
gluon distribution function $f_g(x;l_T^2)$.

After integrating over the azimuth angle of $\vec{l}_T$, the integration
${\cal T}^{(\pm)}$ will then be
\begin{equation}
{\cal T}^{(\pm)}=\pi\frac{\vec{e}^{(\pm)}\cdot \vec{k}_T}{k_T^2}{\cal I},
\end{equation}
where
\begin{eqnarray}
\label{inti}
\nonumber
{\cal I}&=&\int\frac{dl_T^2}{(l_T^2)^2}f_g(x;l_T^2)[\frac{1}{36}(\frac{1}{2}-\frac{k_T^2-l_T^2}{2|k_T^2-l_T^2|})
        +\frac{1+\alpha_k}{4}(\frac{1}{2}-\frac{k_T^2-(1+\alpha_k)l_T^2}{2|k_T^2-(1+\alpha_k)l_T^2|})\\
        &&-\frac{\alpha_k}{4}(\frac{1}{2}-\frac{k_T^2-\alpha_k l_T^2}{2|k_T^2-\alpha_k l_T^2|})].
\end{eqnarray}
Comparing the above results with those of the photoproduction process $\gamma p\rightarrow q\bar q p$\cite{th3,zaka},
we find that the amplitude formula for the diffractive light quark jet hadroproduction 
process $gp\rightarrow q\bar qp$ is much more complicated.
However, the basic structure of the amplitude, especially the expression for the
integration ${\cal I}$ is similar to that for the photoproduction process.
In the integration of (\ref{inti}), if $l_T^2<k_T^2$ the first term of the
integration over $l_T^2$ will be zero; if $l_T^2<k_T^2/(1+\alpha_k)^2$ the second
term will be zero; if $l_T^2<k_T^2/\alpha_k^2$ the third term will be zero.
So, the dominant regions contributing to the three integration terms are
$l_T^2\sim k_T^2$, $l_T^2\sim k_T^2/(1+\alpha_k)^2$, and $l_T^2\sim k_T^2/\alpha_k^2$
respectively.
Approximately, by ignoring some evolution effects of the differential gluon
distribution function $f_g(x;l_T^2)$ in the above dominant integration regions,
we get the following results for the integration ${\cal I}$,
\begin{equation}
\label{i1}
{\cal I}=\frac{1}{k_T^2}[\frac{1}{36}f_g(x;k_T^2)+\frac{(1+\alpha_k)^3}{4}f_g(x;\frac{k_T^2}{(1+\alpha_k)^2})
        -\frac{\alpha_k^3}{4}f_g(x;\frac{k_T^2}{\alpha_k^2})].
\end{equation}

Obtained the formula for the integration ${\cal I}$, the amplitude squared for
the partonic process $gp\rightarrow q\bar qp$ will be reduced to, after averaging
over the spin and color degrees of freedom,
\begin{equation}
\overline{|{\cal A}|}^2=\frac{9}{4}\alpha_s^3(4\pi)^3\pi^2s^2\frac{|{\cal I}|^2}{M_X^2}
        (1-\frac{2k_T^2}{M_X^2}).
\end{equation}
And the cross section for the partonic process $gp\rightarrow q\bar qp$ is
\begin{equation}
\label{xsp}
\frac{d\hat\sigma(gp\rightarrow q\bar qp)}{dt}|_{t=0}=\int_{M_X^2>4k_T^2}dM_X^2dk_T^2
        \frac{9\alpha_s^3\pi^2}{8(M_X^2)^2}\frac{1}{\sqrt{1-\frac{4k_T^2}{M_X^2}}}
        (1-\frac{2k_T^2}{M_X^2})|{\cal I}|^2.
\end{equation}
The integral bound $M_X^2>4k_T^2$ above shows that the dominant contribution
of the integration over $M_X^2$ comes from the region of $M_X^2\sim 4k_T^2$.
Using Eq.~(\ref{ak}), this indicates that in this dominant region 
$\alpha_k$ is of order of 1.
So, in the integration ${\cal I}$ the differential gluon distribution
function $f_g(x;Q^2)$ of the three terms can approximately take their values
at the same scale of $Q^2=k_T^2$.
That is, the integration ${\cal I}$ is then simplified to
\begin{equation}
\label{i2}
{\cal I}\approx \frac{10M_X^2-27k_T^2}{36M_X^2}f_g(x;k_T^2).
\end{equation}
Numerical calculations show that there is little difference (within $10\%$
for $k_T>5~GeV$) between the cross sections by using these two different
parametrizations of ${\cal I}$, Eq.~(\ref{i1}) and Eq.~(\ref{i2}).
So, in Sec.III, we use Eqs.~(\ref{xsp}) and (\ref{i2}) to
estimate the diffractive light quark jet production rate at the Fermilab
Tevatron.

\subsection{Recalculate the heavy quark jet production using the helicity amplitude method}

For a crossing check, in this subsection we will recalculate the diffractive
heavy quark jet production at hadron colliders by using the helicity amplitude
method. In Ref.\cite{charm}, we have calculated this process in the leading
logarithmic approximation of QCD, where we expanded the amplitude in terms of
$l_T^2$. Now, if we use the helicity amplitude method, we donot need to use
the expansion method for the $\Gamma_\mu$ factor in Eq.~(\ref{ima}) as in \cite{charm}.
We can firstly calculate the amplitude explicitly by using the helicity amplitude
method.

However, for the massive fermion, the amplitude formula is more complicated.
Following Ref.\cite{ham1}, we first define the basic spinors $u_\pm(k_0)$ as,
\begin{eqnarray}
\label{k1}
\nonumber
u_+(k_0)=\not\! k_1u_-(k_0),\\
u_-(k_0)\bar u_-(k_0)=\frac{1}{2}(1-\gamma_5)\not\! k_0,
\end{eqnarray}
where the momenta $k_0$ and $k_1$ satisfy the following relations,
\begin{equation}
\label{k2}
k_0\cdot k_0=0,~~~k_1\cdot k_1=-1,~~~k_0\cdot k_1=0.
\end{equation}
Using Eqs.~(\ref{k1}) and (\ref{k2}), we can easily find that the spinor $u_+(k_0)$
satisfies
\begin{equation}
u_+(k_0)\bar u_+(k_0)=\frac{1}{2}(1+\gamma_5)\not\! k_0.
\end{equation}
Provided the basic spinors, we then express any spinors $u(p_i)$ in terms of the basic
ones,\cite{ham1}
\begin{equation}
u_\pm(p_i)=\frac{(\not\! p_i +m_i)u_\pm(k_0)}{\sqrt{2p_i\cdot k_0}}.
\end{equation}
It is easily checked that these spinors satisfy Dirac's equations.
Now, for the massive fermions, the helicity amplitudes for the processes involving
Dirac particles have the following general forms,
\begin{eqnarray}
\label{ham1}
\nonumber
\bar u_+(p_f)Qv_-(p_i)&=&\frac{Tr[Q(\not\! p_i-m_i)(1-\gamma_5)\not\! k_0(\not\! p_f+m_f)]}
        {4\sqrt{(k_0\cdot p_i)(k_0\cdot p_f)}},\\
\nonumber
\bar u_-(p_f)Qv_+(p_i)&=&\frac{Tr[Q(\not\! p_i-m_i)(1+\gamma_5)\not\! k_0(\not\! p_f+m_f)]}
        {4\sqrt{(k_0\cdot p_i)(k_0\cdot p_f)}},\\
\nonumber
\bar u_+(p_f)Qv_+(p_i)&=&\frac{Tr[Q(\not\! p_i-m_i)(1-\gamma_5)\not\! k_1\not\! k_0(\not\! p_f+m_f)]}
        {4\sqrt{(k_0\cdot p_i)(k_0\cdot p_f)}},\\
\bar u_-(p_f)Qv_-(p_i)&=&\frac{Tr[Q(\not\! p_i-m_i)\not\! k_1(1-\gamma_5)\not\! k_0(\not\! p_f+m_f)]}
        {4\sqrt{(k_0\cdot p_i)(k_0\cdot p_f)}},
\end{eqnarray}
where $m_i$ and $m_f$ are the masses for the momenta $p_i$ and $p_f$ respectively,
where $p_i^2=m_i^2,~p_f^2=m_f^2$.
From the above equations, we can see that for the massless fermions ($m_i=m_f=0$)
the formula Eq.~(\ref{ham1}) will then turn back to the formula Eq.~(\ref{ham}).

To calculate the imaginary part of the amplitude Eq.~(\ref{ima}) for the partonic process
$gp\to c\bar c p$, a convenient choice
for the momenta $k_0$ and $k_1$ is,
\begin{equation}
k_0=p,~~~~k_1=e,
\end{equation}
where the vector $e$ is the polarization vector for the incident gluon defined
in Eq.~(\ref{ev}).
Using the formula Eq.~(\ref{ham1}) and the above choice for the momenta $k_0$ and $k_1$,
the helicity amplitudes for Eq.~(\ref{ima}) will then be,
\begin{eqnarray}
\nonumber
{\rm Im}{\cal A}(\pm,+,-)&=&\alpha_k^2(1+\alpha_k){\cal N}\times {\cal T}_c^{(\pm)},\\
{\rm Im}{\cal A}(\pm,-,+)&=&\alpha_k(1+\alpha_k)^2{\cal N}\times {\cal T}_c^{(\pm)},\\
{\rm Im}{\cal A}(\pm,+,+)&=&{\rm Im}{\cal A}(\pm,-,-)=\alpha_k(1+\alpha_k){\cal N}\times \frac{\pi m_c}{2}{\cal I}'_c,
\end{eqnarray}
where ${\cal N}$ is the same as in Eq.~(\ref{ne}), and the integrations ${\cal T}_c^{(\pm)}$ and ${\cal I}'_c$
are defined as 
\begin{eqnarray}
\label{int3}
\nonumber
{\cal T}_c^{(\pm)}&=&\int\frac{d^2\vec{l}_T}{(l_T^2)^2}f(x',x'';l_T^2)[(\frac{1}{2}-\frac{2}{9})
        \frac{\vec{e}^{(\pm)}\cdot \vec{k}_T}{k_T^2+m_c^2}-\frac{1}{36}\frac{\vec{e}^{(\pm)}\cdot(\vec{k}_T-\vec{l}_T)}{m_c^2+(\vec{k}_T-\vec{l}_T)^2}\\
&&-\frac{1+\alpha_k}{4}\frac{\vec{e}^{(\pm)}\cdot(\vec{k}_T-(1+\alpha_k)\vec{l}_T)}{m_c^2+(\vec{k}_T-(1+\alpha_k)\vec{l}_T)^2}
  +\frac{\alpha_k}{4}\frac{\vec{e}^{(\pm)}\cdot(\vec{k}_T-\alpha_k\vec{l}_T)}{m_c^2+(\vec{k}_T-\alpha_k\vec{l}_T)^2}],\\
\nonumber
{\cal I}'_c&=&\frac{1}{\pi}\int\frac{d^2\vec{l}_T}{(l_T^2)^2}f(x',x'';l_T^2)[(\frac{1}{2}-\frac{2}{9})
        \frac{1}{k_T^2+m_c^2}-\frac{1}{36}\frac{1}{m_c^2+(\vec{k}_T-\vec{l}_T)^2}\\
&&-\frac{1+\alpha_k}{4}\frac{1}{m_c^2+(\vec{k}_T-(1+\alpha_k)\vec{l}_T)^2}
  +\frac{\alpha_k}{4}\frac{1}{m_c^2+(\vec{k}_T-\alpha_k\vec{l}_T)^2}].
\end{eqnarray}
If we approximate the differential off-diagonal gluon
distribution function $f(x',x'';l_T^2)$ by the usual diagonal differential
gluon distribution function $f_g(x;l_T^2)$, the above integrations will then
be reduced to, after integrating over the azimuth angle of $\vec{l}_T$,
\begin{equation}
{\cal T}_c^{(\pm)}=\pi\vec{e}^{(\pm)}\cdot \vec{k}_T{\cal I}_c,
\end{equation}
where
\begin{eqnarray}
\label{int4}
\nonumber
{\cal I}_c&=&\int\frac{d^2\vec{l}_T}{(l_T^2)^2}f_g(x;l_T^2)[\frac{5}{18}
        \frac{1}{m_T^2}-\frac{5}{36}\frac{1}{k_T^2}-\frac{1}{72}\frac{k_T^2-m_c^2-l_T^2}{m_1^2}\\
        &&-\frac{1+\alpha_k}{8}\frac{k_T^2-m_c^2-(1+\alpha_k)^2l_T^2}{m_2^2} +\frac{\alpha_k}{8}\frac{k_T^2-m_c^2-\alpha_k^2l_T^2}{m_3^2}],
\end{eqnarray}
where
\begin{eqnarray}
\nonumber
m_T^2&=&k_T^2+m_c^2,~~m_1^2=\sqrt{(m_T^2+l_T^2)^2-k_T^2l_T^2},~~m_2^2=\sqrt{(m_T^2+(1+\alpha_k)^2l_T^2)^2-(1+\alpha_k)^2k_T^2l_T^2},\\
        ~~m_3^2&=&\sqrt{(m_T^2+\alpha_k^2l_T^2)^2-\alpha_k^2k_T^2l_T^2},
\end{eqnarray}
and
\begin{equation}
\label{int5}
{\cal I}'_c=\int\frac{d^2\vec{l}_T}{(l_T^2)^2}f_g(x;l_T^2)[\frac{5}{18}
        \frac{1}{m_T^2}-\frac{1}{36}\frac{1}{m_1^2}-
        \frac{1+\alpha_k}{4}\frac{1}{m_2^2} +\frac{\alpha_k}{4}\frac{1}{m_3^2}].
\end{equation}
So, the amplitude squared for
the partonic process $gp\rightarrow c\bar cp$ will then be reduced to, after averaging
over the spin and color degrees of freedom,
\begin{equation}
\label{cam}
\overline{|{\cal A}|}^2=\frac{9}{4}\alpha_s^3(4\pi)^3\pi^2s^2\frac{m_T^2}{M_X^2}
        [(1-\frac{2k_T^2}{M_X^2})k_T^2|{\cal I}_c|^2+m_c^2|{\cal I}'_c|^2].
\end{equation}

From the above results, we can see that the integrals of Eqs.~(\ref{int4}) and
(\ref{int5}) are proportional to $1/l_T^2$ in the limit of $l_T^2\rightarrow 0$.
That is to say that there exist large logarithmic contributions from the integration region
of $1/R^2_N\ll l_T^2\ll m_T^2$ for the integration over $l_T^2$ as in Ref.\cite{charm}.
So, we can expand the integrals of Eqs.~(\ref{int4}) and (\ref{int5}) in terms
of $l_T^2$ to get the leading logarithmic contribution to the amplitude.
In the limit of $l_T^2\rightarrow 0$, the parameters $m_1^2$, $m_2^2$ and $m_3^2$
scale as,
\begin{eqnarray}
\nonumber
\frac{1}{m_1^2}&\approx & \frac{1}{m_T^2}[1-\frac{m_c^2-k_T^2}{m_T^2}\frac{l_T^2}{m_T^2}],\\
\nonumber
\frac{1}{m_2^2}&\approx & \frac{1}{m_T^2}[1-\frac{m_c^2-k_T^2}{m_T^2}\frac{(1+\alpha_k)^2l_T^2}{m_T^2}],\\
\frac{1}{m_3^2}&\approx & \frac{1}{m_T^2}[1-\frac{m_c^2-k_T^2}{m_T^2}\frac{\alpha_k^2l_T^2}{m_T^2}].
\end{eqnarray}
Under this approximation, the integrations Eqs.~(\ref{int4}) and (\ref{int5}) will then 
be related to the integrated gluon distribution function $xg(x;Q^2)$,
\begin{eqnarray}
{\cal I}_c\approx \frac{2m_c^2}{36(m_T^2)^3M_X^2}(10M_X^2-27m_T^2)xg(x;m_T^2),\\
{\cal I'}_c\approx \frac{m_c^2-k_T^2}{36(m_T^2)^3M_X^2}(10M_X^2-27m_T^2)xg(x;m_T^2).
\end{eqnarray}
Substituting the above results into Eq.~(\ref{cam}), we can then reproduce the leading
logarithmic approximation result for the diffractive charm jet production process
at hadron colliders which has been calculated in\cite{charm}.

\subsection{$qp\rightarrow qgp$ process}

We first give the Sudakov parameters for the loop mentum $l$
for this process (Fig.~3),
\begin{eqnarray}
\nonumber
\alpha_l&=&-\frac{l_T^2}{s},\\
\nonumber
\beta_l&=&\frac{2(k_T,l_T)-l_T^2}{\alpha_ks},~~~{\rm for~Diag.}1,~2,~6,\\
\nonumber
&=&\frac{2(k_T,l_T)+l_T^2}{(1+\alpha_k)s},~~~{\rm for~Diag.}5,~7,~8,\\
&=&-\frac{M_X^2-l_T^2}{s},~~~~~~~{\rm for~Diag.}3,~4,~9.
\end{eqnarray}
And the imaginary part of the amplitude ${\cal A}(qp\rightarrow qgp)$ for each diagram
of Fig.~3 has the following general form,
\begin{equation}
\label{qima}
{\rm Im}{\cal A}=C_F(T_{ij}^a)\int \frac{d^2l_T}{(l_T^2)^2}F\times\bar u
        _i(u-k)\Gamma_\mu u_j(q),
\end{equation}
where $C_F$ is the color factor for each diagram.
$a$ is the color index of the incident gluon.
$\Gamma_\mu$ represents some $\gamma$ matrices including one
propagator. $F$ is the same as in Eq.~(\ref{feq}).

In \cite{photon}, we calculated the diffractive photon
production process $qp\rightarrow \gamma qp$, in which the Feynman diagrams
are similar to the first four diagrams of Fig.~3.
In that paper, we calculate the cross section by directly squaring the
partonic process amplitude.
However, in the calculations here for the partonic process $qp\rightarrow qgp$
because there are additional five diagrams contribution, it is not convenient
to directly square the amplitude.
Following the above subsections, we calculate the amplitude
by employing the helicity amplitude method.
Furthermore, we will show that by using the helicity amplitude method we can
reproduce the cross section formula for the diffractive photon production
process\cite{photon}.

For the massless quark spinors, we use the definition of Eq.~(\ref{spinor}).
For the polarization vector of the outgoing gluon (its momentum is $k+q$), 
following the method of Ref.\cite{wu}, we find that it is convenient to
choose
\begin{equation}
\label{epq}
\not\! e^{(\pm)}=N_e[(\not\! k+\not\! q)\not\! q\not\! p(1\mp\gamma_5)+
        \not\! p\not\! q(\not\! k+\not\! q)(1\pm\gamma_5)].
\end{equation}
The normalization factor $N_e$ equals to
\begin{equation}
\label{ene}
N_e=\frac{1}{s\sqrt{2k_T^2}}.
\end{equation}
With this definition (\ref{epq}), we can easily get the scalar products between the
four-momenta and the polarization vector $e$ as
\begin{equation}
\label{eprc}
e\cdot p=0,~~e\cdot q=N_e\frac{k_T^2s}{1+\alpha_k},~~e\cdot k_T=-N_ek_T^2s,~~e\cdot l_T=-N_e(k_T,l_T)s.
\end{equation}

And then the helicity amplitudes have the following general forms\cite{ham},
\begin{equation}
\label{ham1p}
\bar u_\pm(p_f)Qu_\pm(p_i)=\frac{Tr[Q\not\! p_i\not\! n\not\! p_f(1\mp\gamma_5)]}
        {4\sqrt{(n\cdot p_i)(n\cdot p_f)}},
\end{equation}
where $n$ is an arbitrary massless 4-vector, which is also set to be $n=p$.
Using this formula (\ref{ham1p}), we can calculate the helicity amplitude
${\cal A}(\lambda_1,\lambda_2,\lambda_3)$ for the diffractive
process $qp\rightarrow qgp$.
Here $\lambda_1$ represents the helicity of the incident quark;
$\lambda_2$ and $\lambda_3$ represent the helicities of the
outgoing gluon and quark respectively.

For the first four diagrams, to sum up together, the imaginary part of the
amplitude ${\cal A}(+,+,+)$ is
\begin{equation}
\label{qim1}
{\rm Im}{\cal A}^{1234}(+,+,+)=\alpha_k^2(1+\alpha_k){\cal N}_q\times
        \int\frac{d^2\vec{l}_T}{(l_T^2)^2}f(x',x'';l_T^2)
        (\frac{2}{9}-\frac{-1}{36}\frac{k_T^2-(1+\alpha_k)(k_T,l_T)}{(\vec{k}_T-(1+\alpha_k)\vec{l}_T)^2}),
\end{equation}
where $\frac{2}{9}$ and $\frac{-1}{36}$ are the color factors for Diags.1,4 and
Diags.2,3 respectively, and
${\cal N}_q$ is defined as
\begin{equation}
{\cal N}_q=\frac{3s}{\sqrt{-2\alpha_kk_T^2}}g_s^3T_{ij}^a.
\end{equation}
The other helicity amplitudes for the first four diagrams have the similar forms
as (\ref{qim1}),
\begin{eqnarray}
\label{im11}
\nonumber
{\rm Im}{\cal A}^{1234}(-,-,-)&=&{\rm Im}{\cal A}^{1234}(+,+,+),\\
{\rm Im}{\cal A}^{1234}(+,-,+)&=&{\rm Im}{\cal A}^{1234}(-,+,-)=\frac{-1}{\alpha_k}{\rm Im}{\cal A}^{1234}(+,+,+).
\end{eqnarray}

These amplitude expressions Eq.~(\ref{qim1}) can also serve as the calculations
of the amplitude for the diffractive direct photon
production process $q p\rightarrow q\gamma p$ \cite{photon} except the difference
on the color factors.\footnote{In Ref.\cite{photon}, we did not employ the
helicity amplitude method. If we use the amplitude expressions Eqs.(\ref{qim1}) and
(\ref{im11}) (correct the color factors) to calculate the photon production process $qp\rightarrow q\gamma p$,
we can get the same result as that in \cite{photon}. This can be viewed as a cross
check for the methods we used in the calculations.}
In the direct photon process, the color
factors for these four diagrams are the same (they are all $\frac{2}{9}$).

As discussed in the above subsection of the calculation for the
partonic processes $gp\rightarrow q\bar qp$, we must first exam the amplitude
(\ref{qima}) behavor under the limit of $l_T^2\rightarrow 0$ in the integration over $l$
to avoid the linear singularities.
From Eq.~(\ref{qim1}), we can see that the amplitude for the
diffractive direct photon production process $qp\rightarrow q\gamma p$ is
exact zero at $l_T^2\rightarrow 0$.
However, for the process $qp\rightarrow qgp$ the amplitude
for the first four diagrams 
is not exact zero in the limit $l_T^2\rightarrow 0$ due to the inequality of the
color factors between them.
That is to say, the sum of the first four diagrams is not free of linear
singularities.
So, the contributions from the last five diagrams of Fig.~3 must
cancel out the linear singularity which rises from the first four diagrams
to guarantee the total sum of the amplitude from all of the diagrams free
of linear singularities.

Finally, by adding up all of the nine diagrams of Fig.2, the 
imaginary parts of the amplitudes
are
\begin{eqnarray}
\nonumber
{\rm Im}{\cal A}(+,+,+)&=&{\rm Im}{\cal A}(-,-,-)=\frac{\alpha_k^2}{4}{\cal N}_q\times {\cal T},\\
{\rm Im}{\cal A}(+,-,+)&=&{\rm Im}{\cal A}(-,+,-)=-\frac{\alpha_k}{4}{\cal N}_q\times {\cal T},
\end{eqnarray}
where
\begin{eqnarray}
\label{qint2}
\nonumber
{\cal T}&=&\int\frac{d^2\vec{l}_T}{(l_T^2)^2}f(x',x'';l_T^2)[
        \frac{(1+\alpha_k)^2}{9}\frac{(k_T,l_T)-(1+\alpha_k)l_T^2}{(\vec{k}_T-(1+\alpha_k)\vec{l}_T)^2}
        -(1+\alpha_k)\frac{(k_T,l_T)+l_T^2}{(\vec{k}_T+\vec{l}_T)^2}\\
  &&-\alpha_k\frac{(k_T,l_T)-l_T^2}{(\vec{k}_T-\vec{l}_T)^2}
  +\alpha_k^2\frac{(k_T,l_T)-\alpha_kl_T^2}{(\vec{k}_T-\alpha_k\vec{l}_T)^2}].
\end{eqnarray}
From the above results, we can see that in the integration of the amplitude
the linear singularity from different diagrams are canceled out by each other,
which will guarantee there is no linear singularity in the total sum.

Following the argument in the above subsection of the calculation for the partonic
process $gp\rightarrow q\bar qp$, we can also approximate the differential off-diagonal gluon
distribution function $f(x',x'';l_T^2)$ by the usual diagonal differential
gluon distribution function $f_g(x;l_T^2)$ in the integration ${\cal T}$, by which 
we can further simplify the integration of ${\cal T}$.
After integrated over the azimuth angle of $\vec{l}_T$, the integration
${\cal T}$ will then be
\begin{eqnarray}
\label{qt}
\nonumber
{\cal T}&=&\pi\int\frac{dl_T^2}{(l_T^2)^2}f_g(x;l_T^2)[
        \frac{1+\alpha_k}{9}(\frac{1}{2}-\frac{k_T^2-(1+\alpha_k)l_T^2}{2|k_T^2-(1+\alpha_k)l_T^2|})
        +(\frac{1}{2}-\frac{k_T^2-l_T^2}{2|k_T^2-l_T^2|})\\
        &&+\alpha_k(\frac{1}{2}-\frac{k_T^2-\alpha_k l_T^2}{2|k_T^2-\alpha_k l_T^2|})].
\end{eqnarray}
In the above integration, if $l_T^2<k_T^2/(1+\alpha_k)^2$ the first term of the
integration over $l_T^2$ will be zero; if $l_T^2<k_T^2$ the second
term will be zero; if $l_T^2<k_T^2/\alpha_k^2$ the third term will be zero.
So, the dominant regions contributing to the three integration terms are
$l_T^2\sim k_T^2/(1+\alpha_k)^2$, $l_T^2\sim k_T^2$, and $l_T^2\sim k_T^2/\alpha_k^2$
respectively.
Approximately, by ignoring some evolution effects of the differential gluon
distribution function $f_g(x;l_T^2)$ in the above dominant integration regions,
we get the following results for the integration ${\cal T}$,
\begin{equation}
\label{qi1}
{\cal T}=\frac{\pi}{k_T^2}[f_g(x;k_T^2)+\frac{(1+\alpha_k)^3}{9}f_g(x;\frac{k_T^2}{(1+\alpha_k)^2})
        +\alpha_k^3f_g(x;\frac{k_T^2}{\alpha_k^2})].
\end{equation}

Obtained the formula for the integration ${\cal T}$, the amplitude squared for
the partonic process $qp\rightarrow qgp$ will be reduced to, after averaging
over the spin and color degrees of freedom,
\begin{equation}
\overline{|{\cal A}|}^2=\frac{\alpha_s^3(4\pi)^3}{24}\frac{1+\alpha_k^2}{M_X^2(1+\alpha_k)}s^2|{\cal T}|^2.
\end{equation}
And the cross section for the partonic process $qp\rightarrow qgp$ is
\begin{eqnarray}
\label{qxsp}
\nonumber
\frac{d\hat\sigma(qp\rightarrow qgp)}{dt}|_{t=0}&=&\int_{M_X^2>4k_T^2}dM_X^2dk_T^2d\alpha_k[\delta(\alpha_k-\alpha_1)+\delta(\alpha_k-\alpha_2)]\\
        &&\frac{\alpha_s^3}{96(M_X^2)^2}\frac{1+\alpha_k^2}{1+\alpha_k}\frac{1}{\sqrt{1-\frac{4k_T^2}{M_X^2}}}
        |{\cal T}|^2,
\end{eqnarray}
where $\alpha_{1,2}$ are the solutions of the following equations,
\begin{equation}
\alpha(1+\alpha)+\frac{k_T^2}{M_X^2}=0.
\end{equation}
The integral bound $M_X^2>4k_T^2$ in (\ref{qxsp}) shows that the dominant contribution
of the integration over $M_X^2$ comes from the region of $M_X^2\sim 4k_T^2$.
Using Eq.~(\ref{ak}), this indicates that in this dominant region 
$\alpha_k$ is of order of 1.
So, in the integration ${\cal T}$ the differential gluon distribution
function $f_g(x;Q^2)$ of the three terms can approximately take their values
at the same scale of $Q^2=k_T^2$.
That is, the integration ${\cal T}$ is then simplified to
\begin{equation}
\label{qi2}
{\cal T}=\frac{\pi}{9k_T^2}f_g(x;k_T^2)(1+\alpha_k)(10-7\alpha_k+10\alpha_k^2).
\end{equation}

\subsection{$gp\rightarrow ggp$ process}

For the partonic process $gp\rightarrow ggp$, there are twelve diagrams in the
leading order contributions as shown in Fig.~4.
The first nine diagrams are due to the existence of the three-gluon interaction vertex, and
the last three diagrams are due to the existence of  the four-gluon interaction vertex.
But it will be shown in the following calculations, the last three diagrams
do not contribute under some choice of the polarizations of the three
external gluons.

As usual, we first give the Sudakov parameters for the loop momentum $l$,
\begin{eqnarray}
\nonumber
\alpha_l&=&-\frac{l_T^2}{s},\\
\nonumber
\beta_l&=&\frac{2(k_T,l_T)-l_T^2}{\alpha_ks},~~~{\rm for~Diag.}1,~4,~6,~10,\\
\nonumber
&=&\frac{2(k_T,l_T)+l_T^2}{(1+\alpha_k)s},~~~{\rm for~Diag.}2,~3,~5,~11,\\
&=&-\frac{M_X^2-l_T^2}{s},~~~~~~~{\rm for~Diag.}7,~8,~9,~12.
\end{eqnarray}
And the imaginary part of the amplitude ${\cal A}(gp\rightarrow ggp)$ for each diagram
of Fig.~4 has the following general form,
\begin{equation}
\label{gima}
{\rm Im}{\cal A}=C_Ff_{abc}\int \frac{d^2l_T}{(l_T^2)^2}G(e_1,e_2,e_3)\times F,
\end{equation}
where $C_F$ is the color factor for each diagram.
$a,~b,~c$ are the color indexes for the incident gluon and the two outgoing
gluons respectively, and $f_{abc}$ are the antisymmetric $SU(3)$ structure
constants.
$G(e_1,e_2,e_3)$ represents the interaction part including one propagator
for the first nine diagrams, where
$e_1,~e_2,~e_3$ are the polarization vectors for the incident gluon
and the two outgoing gluons.
$F$ in the integral is the same as that in Eq.(\ref{feq}).

The color factors $C_F$ for the twelve diagrams are 
\begin{eqnarray}
\label{cf}
\nonumber
C_F&=&\frac{1}{2},~~~~~~~{\rm for~ Diag.}1,~7,\\
\nonumber
C_F&=&-\frac{1}{2},~~~~~{\rm for~ Diag.}2,\\
\nonumber
C_F&=&-\frac{1}{4},~~~~~{\rm for~ Diag.}3,~6,~9,\\
\nonumber
C_F&=&\frac{1}{4},~~~~~~~{\rm for~ Diag.}4,~5,~8,\\
C_F&=&\frac{3}{4},~~~~~~~{\rm for~ Diag.}10,~11,~12.
\end{eqnarray}

Following the calculation method used in the above subsections, we employ the helicity amplitude method
to calculate the amplitude Eq.(\ref{gima}).
For the polarization vector of the incident gluon, which is transversely polarized,
we use the same definition in Eq.~(\ref{ev}).
For the two outgoing gluons, we choose their polarization vectors as\cite{wu}
\begin{eqnarray}
\label{geprc}
\nonumber
\not\! e_2^{(\pm)}=N_e[(\not\! k+\not\! q)\not\! q\not\! p(1\mp\gamma_5)+
        \not\! p\not\! q(\not\! k+\not\! q)(1\pm\gamma_5)],\\
\not\! e_3^{(\pm)}=N_e[(\not\! u-\not\! k)\not\! q\not\! p(1\mp\gamma_5)+
        \not\! p\not\! q(\not\! u-\not\! k)(1\pm\gamma_5)].
\end{eqnarray}
The normalization factor $N_e$ has the same form as in Eq.(\ref{ene}).
Under the above choice of the polarization vectors for the external gluons,
we can easily find that they are satisfied the following equations,
\begin{equation}
p\cdot e_1=p\cdot e_2=p\cdot e_3=0.
\end{equation}
With these relations, we can further find that the last three diagrams do
not contribute to the partonic process $gp\rightarrow ggp$.

For the first nine diagrams, there are two helicity amplitudes among
the eight helicity amplitudes do not contribute in the context of the above choice
of the polarizations of the external gluons, i.e.,
\begin{equation}
{\rm Im}{\cal A}(+,+,+)={\rm Im}{\cal A}(-,-,-)=0.
\end{equation}
In the expression of the amplitude ${\cal A}(\lambda(e_1),\lambda(e_2),\lambda(e_3))$,
$\lambda$ denote the helicities for the three gluons respectively.
The other six helicity amplitudes are divided into the following three different
sets,
\begin{eqnarray}
\nonumber
{\rm Im}{\cal A}(+,-,-)&\sim &{\rm Im}{\cal A}(-,+,+),\\
\nonumber
{\rm Im}{\cal A}(+,-,+)&\sim &{\rm Im}{\cal A}(-,+,-),\\
{\rm Im}{\cal A}(+,+,-)&\sim &{\rm Im}{\cal A}(-,-,+).
\end{eqnarray}

For the first helicity amplitudes set, ${\rm Im}{\cal A}(\pm,\mp,\mp)$,
to sum up all of the nine diagrams, we get
\begin{equation}
\label{gm1}
{\rm Im}{\cal A}(\pm,\mp,\mp)={\cal N}'\pi \vec{e}_1^{(\pm)}\cdot k_T{\cal I}_g,
\end{equation}
where ${\cal N}'$ is defined as
\begin{equation}
{\cal N}'=\frac{3}{4}\frac{s}{k_T^2}g_s^3f_{abc}.
\end{equation}
And the integration ${\cal I}_g$ is
\begin{eqnarray}
\label{gim11}
\nonumber
{\cal I}_g&=&{1\over \pi}\int\frac{d^2\vec{l}_T}{(l_T^2)^2}f(x',x'';l_T^2)[-(1+\alpha_k)\frac{k_T^2+(k_T,l_T)}{(\vec{k}_T+\vec{l}_T)^2}
        +\alpha_k\frac{k_T^2-(k_T,l_T)}{(\vec{k}_T-\vec{l}_T)^2}
        \\
  &&+(1+\alpha_k)^2\frac{k_T^2-(1+\alpha_k)(k_T,l_T)}{(\vec{k}_T-(1+\alpha_k)\vec{l}_T)^2}
  -\alpha_k^2\frac{k_T^2-\alpha_k(k_T,l_T)}{(\vec{k}_T-\alpha_k\vec{l}_T)^2}]\\
\nonumber
&=&{1\over \pi}\int\frac{d^2\vec{l}_T}{(l_T^2)^2}f(x',x'';l_T^2)[(1+\alpha_k)\frac{(k_T,l_T)+l_T^2}{(\vec{k}_T+\vec{l}_T)^2}
        +\alpha_k\frac{(k_T,l_T)-l_T^2}{(\vec{k}_T-\vec{l}_T)^2}        
        \\
  &&-(1+\alpha_k)^2\frac{(k_T,l_T)-(1+\alpha_k)l_T^2}{(\vec{k}_T-(1+\alpha_k)\vec{l}_T)^2}
  +\alpha_k^2\frac{(k_T,l_T)-\alpha_kl_T^2}{(\vec{k}_T-\alpha_k\vec{l}_T)^2}].
\end{eqnarray}
From the above equations, we can check that there is no linear singularity at the
limit of $l_T^2\rightarrow 0$ in the integration of the amplitude over
the loop momentum.
The first term of the integration ${\cal I}_g$ in Eq.(\ref{gim11}) comes from the
contribution of Diag.3; the second term comes from Diag.4; the third term comes from Diag.6 and Diag.9;
the last term comes from Diag.5 and Diag.8. The contributions from Diags.1, 2
and 7 are canceled out by each other.
From (\ref{gim11}), we can see that the linear singularities coming from the
four terms are canceled out by each other.
The final result for the amplitude is now free of linear singularity.
We must emphasize here that only the total sum of the contributions from  all
of the diagrams is free of linear singularity. The separation of these diagrams
will cause linear singularity.

Using the same approximations used in the above subsections,
i.e., approximating the differential off-diagonal gluon
distribution function $f(x',x'';l_T^2)$ by the usual diagonal differential
gluon distribution function $f_g(x;l_T^2)$, we can further simplify the integration
of ${\cal I}_g$.
After integrated over the azimuth angle of $\vec{l}_T$, this integration
will then be
\begin{eqnarray}
\label{gi2}
\nonumber
{\cal I}_g&=&\int\frac{dl_T^2}{(l_T^2)^2}f_g(x;l_T^2)[(\frac{1}{2}-\frac{k_T^2-l_T^2}{2|k_T^2-l_T^2|})
        -{(1+\alpha_k)}(\frac{1}{2}-\frac{k_T^2-(1+\alpha_k)l_T^2}{2|k_T^2-(1+\alpha_k)l_T^2|})
        \\
        &&+\alpha_k(\frac{1}{2}-\frac{k_T^2-\alpha_k l_T^2}{2|k_T^2-\alpha_k l_T^2|})].
\end{eqnarray}
The above equation shows that the integration ${\cal I}_g$ here has the similar
behavior as that of the integration ${\cal T}$ of Eq.(\ref{qt}) in the last
subsection.
So, the three terms of the above integration ${\cal I}_g$ are dominantly 
contributed from the integral regions of $l_T^2$ as
$l_T^2\sim k_T^2/(1+\alpha_k)^2$, $l_T^2\sim k_T^2$, and $l_T^2\sim k_T^2/\alpha_k^2$
respectively.
Approximately, we may also ignore the evolution effects of the differential gluon
distribution function $f_g(x;l_T^2)$ in the above dominant integration regions,
and so the integration ${\cal I}_g$ is reduced to
\begin{equation}
\label{gi1}
{\cal I}_g=\frac{1}{k_T^2}[f_g(x;k_T^2)-(1+\alpha_k)^3f_g(x;\frac{k_T^2}{(1+\alpha_k)^2})
        +\alpha_k^3f_g(x;\frac{k_T^2}{\alpha_k^2})].
\end{equation}

For the second helicity amplitudes set, ${\rm Im}{\cal A}(\pm,\mp,\pm)$,
the calculations are more complicated, and
the contribution from Diag.3 is
\begin{eqnarray}
\label{g21}
\nonumber
{\rm Im}{\cal A}^3(\pm,\mp,\pm)&=&-{\cal N}'(1+\alpha_k)^2\int\frac{d^2\vec{l}_T}{(l_T^2)^2}f(x',x'';l_T^2)\\
       && \frac{\alpha_kk_T^2\vec{e}_1^{(\pm)}\cdot(\vec{k}_T+\vec{l}_T)+
        (k_T^2+(k_T,l_T))\vec{e}_1^{(\pm)}\cdot\vec{k}_T}{(\vec{k}_T+\vec{l}_T)^2}.
\end{eqnarray}
The contribution from Diag.4 is
\begin{eqnarray}
\label{g22}
\nonumber
{\rm Im}{\cal A}^4(\pm,\mp,\pm)&=&{\cal N}'\alpha_k(1+\alpha_k)\int\frac{d^2\vec{l}_T}{(l_T^2)^2}f(x',x'';l_T^2)\\
       && \frac{\alpha_kk_T^2\vec{e}_1^{(\pm)}\cdot(\vec{k}_T-\vec{l}_T)+
        (k_T^2-(k_T,l_T))\vec{e}_1^{(\pm)}\cdot\vec{k}_T}{(\vec{k}_T-\vec{l}_T)^2}.
\end{eqnarray}
The contributions from Diag.5 and Diag.8, to sum up together, are
\begin{eqnarray}
\label{g23}
\nonumber
{\rm Im}{\cal A}^{58}(\pm,\mp,\pm)&=&-{\cal N}'\alpha_k(1+\alpha_k)\int\frac{d^2\vec{l}_T}{(l_T^2)^2}f(x',x'';l_T^2)\\
       && \frac{\alpha_kk_T^2\vec{e}_1^{(\pm)}\cdot(\vec{k}_T-\alpha_k\vec{l}_T)+
        (k_T^2-\alpha_k(k_T,l_T))\vec{e}_1^{(\pm)}\cdot\vec{k}_T}{(\vec{k}_T-\alpha_k\vec{l}_T)^2}.
\end{eqnarray}
The contributions from Diag.6 and Diag.9, to sum up together, are
\begin{eqnarray}
\label{g24}
\nonumber
{\rm Im}{\cal A}^{69}(\pm,\mp,\pm)&=&{\cal N}'(1+\alpha_k)^2\int\frac{d^2\vec{l}_T}{(l_T^2)^2}f(x',x'';l_T^2)\\
       && \frac{\alpha_kk_T^2\vec{e}_1^{(\pm)}\cdot(\vec{k}_T-(1+\alpha_k)\vec{l}_T)+
        (k_T^2-(1+\alpha_k)(k_T,l_T))\vec{e}_1^{(\pm)}\cdot\vec{k}_T}{(\vec{k}_T-(1+\alpha_k)\vec{l}_T)^2}.
\end{eqnarray}
The contributions from other three diagrams (Diag.1, Diag.2 and Diag.7) are
canceled out by each other.
From the above results Eqs.(\ref{g21}-\ref{g24}), we can see that every term
has linear singularity at the limit of $l_T^2\rightarrow 0$ in the integration
of the amplitude over $l_T^2$, while
their total sum is free of the linear singularity.

Following the procedure as we do for the helicity amplitude ${\cal A}(\pm,\mp,\mp)$
in the above, we can approximate the off-diagonal gluon distribution function
$f(x',x'';l_T^2)$ by the usual diagonal differential
gluon distribution function $f_g(x;l_T^2)$.
After integrating over the azimuth angle of $\vec{l}_T$, to sum up all of
Eqs.(\ref{g21}-\ref{g24}), we get the helicity amplitude,
\begin{equation}
\label{gm2}
{\rm Im}{\cal A}(\pm,\mp,\pm)={\cal N}'\pi (1+\alpha_k)^2\vec{e}_1^{(\pm)}\cdot k_T{\cal I}_g,
\end{equation}
where ${\cal I}_g$ is the same as Eq.(\ref{gi2}) and then Eq.(\ref{gi1}) under
the same approximation.

For the third helicity amplitudes set, ${\rm Im}{\cal A}(\pm,\pm,\mp)$,
the calculations are similar to the calculations of ${\rm Im}{\cal A}(\pm,\mp,\pm)$.
The contribution from Diag.3 is
\begin{eqnarray}
\label{g31}
\nonumber
{\rm Im}{\cal A}^3(\pm,\pm,\mp)&=&-{\cal N}'\alpha_k(1+\alpha_k)\int\frac{d^2\vec{l}_T}{(l_T^2)^2}f(x',x'';l_T^2)\\
       && \frac{(1+\alpha_k)k_T^2\vec{e}_1^{(\pm)}\cdot(\vec{k}_T+\vec{l}_T)-
        (k_T^2+(k_T,l_T))\vec{e}_1^{(\pm)}\cdot\vec{k}_T}{(\vec{k}_T+\vec{l}_T)^2}.
\end{eqnarray}
The contribution from Diag.4 is
\begin{eqnarray}
\label{g32}
\nonumber
{\rm Im}{\cal A}^4(\pm,\pm,\mp)&=&{\cal N}'\alpha_k^2\int\frac{d^2\vec{l}_T}{(l_T^2)^2}f(x',x'';l_T^2)\\
       && \frac{(1+\alpha_k)k_T^2\vec{e}_1^{(\pm)}\cdot(\vec{k}_T-\vec{l}_T)-
        (k_T^2-(k_T,l_T))\vec{e}_1^{(\pm)}\cdot\vec{k}_T}{(\vec{k}_T-\vec{l}_T)^2}.
\end{eqnarray}
The contributions from Diag.5 and Diag.8, to sum up together, are
\begin{eqnarray}
\label{g33}
\nonumber
{\rm Im}{\cal A}^{58}(\pm,\pm,\mp)&=&-{\cal N}'\alpha_k^2\int\frac{d^2\vec{l}_T}{(l_T^2)^2}f(x',x'';l_T^2)\\
       && \frac{(1+\alpha_k)k_T^2\vec{e}_1^{(\pm)}\cdot(\vec{k}_T-\alpha_k\vec{l}_T)-
        (k_T^2-\alpha_k(k_T,l_T))\vec{e}_1^{(\pm)}\cdot\vec{k}_T}{(\vec{k}_T-\alpha_k\vec{l}_T)^2}.
\end{eqnarray}
The contributions from Diag.6 and Diag.9, to sum up together, are
\begin{eqnarray}
\label{g34}
\nonumber
{\rm Im}{\cal A}^{69}(\pm,\pm,\mp)&=&{\cal N}'\alpha_k(1+\alpha_k)\int\frac{d^2\vec{l}_T}{(l_T^2)^2}f(x',x'';l_T^2)\\
       && \frac{(1+\alpha_k)k_T^2\vec{e}_1^{(\pm)}\cdot(\vec{k}_T-(1+\alpha_k)\vec{l}_T)-
        (k_T^2-(1+\alpha_k)(k_T,l_T))\vec{e}_1^{(\pm)}\cdot\vec{k}_T}{(\vec{k}_T-(1+\alpha_k)\vec{l}_T)^2}.
\end{eqnarray}
And also, we find that the contributions from other three diagrams (Diag.1, Diag.2 and Diag.7) are
canceled out by each other, and the total sum of 
Eqs.(\ref{g31}-\ref{g34}) is free of the linear singularity.
If we approximate the off-diagonal gluon distribution function
$f(x',x'';l_T^2)$ by the usual diagonal differential
gluon distribution function $f_g(x;l_T^2)$,
and integrate over the azimuth angle of $\vec{l}_T$, their sum will lead to
a similar result as in Eq.(\ref{gm2}),
\begin{equation}
\label{gm3}
{\rm Im}{\cal A}(\pm,\pm,\mp)={\cal N}'\pi \alpha_k^2\vec{e}_1^{(\pm)}\cdot k_T{\cal I}_g.
\end{equation}

By summing up all of the helicity amplitudes Eqs.(\ref{gm1}), (\ref{gm2}) and
{\ref{gm3}), we will get the amplitude squared for 
the partonic process $gp\rightarrow ggp$, after averaging
over the spin and color degrees of freedom,
\begin{equation}
\overline{|{\cal A}|}^2=\frac{27\pi^2\alpha_s^3(4\pi)^3}{16}\frac{s^2}{k_T^2}(1-\frac{k_T^2}{M_X^2})^2|{\cal I}_g|^2.
\end{equation}
And the cross section for the partonic process $gp\rightarrow ggp$ is
\begin{equation}
\label{gxsp}
\frac{d\hat\sigma(gp\rightarrow ggp)}{dt}|_{t=0}=\int_{M_X^2>4k_T^2}dM_X^2dk_T^2
        \frac{27\alpha_s^3\pi^2}{32M_X^2k_T^2}(1-\frac{k_T^2}{M_X^2})^2|{\cal I}_g|^2\frac{1}{\sqrt{1-\frac{4k_T^2}{M_X^2}}},
\end{equation}
Following the same argument in the last subsection for the calculations of the partonic process
$qp\rightarrow qgp$, 
we see that the dominant contribution of the integration over $M_X^2$
comes from the region of $M_X^2\sim 4k_T^2$, where
the differential gluon distribution
function $f_g(x;Q^2)$ of the three terms in the integration ${\cal I}_g$
can approximately take their values
at the same scale of $Q^2=k_T^2$.
That is, the integration ${\cal I}_g$ is then simplified to
\begin{equation}
\label{i22}
{\cal I}_g=\frac{1}{k_T^2}(-3\alpha_k(1+\alpha_k))f_g(x;k_T^2)
=\frac{1}{k_T^2}\frac{3k_T^2}{M_X^2}f_g(x;k_T^2)=\frac{3}{M_X^2}f_g(x;k_T^2).
\end{equation}

\section{Numerical results}

In this section, we give some numerical results for the diffractive light
quark jet and gluon jet production at the Fermilab Tevatron.
We will study the $p_T$ distribution
and $x_1$ distribution of the cross section.
A more thorough phenomenological study, including a comparison
to currently available data at Tevatron on the diffractive dijet production
rate, will be presented elsewhere.
Again, we emphasize that our numerical results are just for the contribution
from the ``Pomeron Fragmentation" region ($M_X^2\sim 4k_T^2$).

Another important effect which will affect the numerical results is
the nonfactorization effect caused by
the spectator interactions in the hard diffractive processes
in hadron collisions mentioned in the section of Introduction.
Here, we use a suppression factor ${\cal F}_S$ to describe this
nonfactorization effect in the hard diffractive processes at hadron
colliders\cite{soper}.
At the Tevatron,
the value of ${\cal F}_S$ may be as small as ${\cal F}_S\approx 0.1$\cite{soper,tev}.
That is to say, the total cross section of the diffractive processes
at the Tevatron may be reduced down by an order of magnitude due to
this nonfactorization effect.
In the following numerical calculations, we adopt this suppression factor value
to evaluate the diffractive production rate of light quark jet at the Fermilab Tevatron.

The cross section formulas for the three partonic processes have been
calculated in the above section. We will use Eqs.~(\ref{xsp}), (\ref{qxsp}) and
(\ref{gxsp}) to evaluate the production rates for these processes.
In our calculations, the scales for the parton distribution functions and
the running coupling constant
are both set to be $Q^2=k_T^2$. For the parton distribution functions,
we choose the GRV NLO set \cite{grv}.

The numerical results of the diffractive light quark jet production
at the Fermilab Tevatron are plotted in Fig.~5 and Fig.~6.
In Fig.~5, we plot the differential cross section $d\sigma/dt|_{t=0}$
as a function of the lower bound of the transverse momentum of the light
quark jet, $k_{T{\rm min}}$. This figure shows that the cross section is
sensitive to the transverse momentum cut $k_{T{\rm min}}$.
The differential cross section decreases over four orders of magnitude
as $k_{T{\rm min}}$ increases from $5~GeV$ to $15~GeV$.

It is interesting to compare the cross section of the diffractive
light quark jet production with that of the diffractive charm quark jet
production\cite{charm}, which
is also shown in Fig.~5. The two curves in this figure show that the production
rates of the light quark jet and the charm quark jet
in the diffractive processes are in the same order of
magnitude.
However, we know that the cross section of diffractive heavy quark jet production
is related to the integrated gluon distribution function, while the cross section
of the light quark jet production is related to the differential
gluon distribution function.

In Fig.~6, we plot the differential cross section $d\sigma/dt|_{t=0}$ as
a function of the lower bound of the momentum fraction of the proton
carried by the incident gluon $x_{1{\rm min}}$,
where we set $k_{T{\rm min}}=5~GeV$. This figure shows that the dominant
contribution comes from the region of $x_1\sim 10^{-2}-10^{-1}$.
This property is the same as that of the diffractive charm jet production
at the Tevatron\cite{charm}.

In Fig.~7, we plot the differential cross section $d\sigma/dt|_{t=0}$
as a function of the lower bound of the transverse momentum of the gluon
jet, $k_{T{\rm min}}$. This figure shows that the cross section is
sensitive to the transverse momentum cut $k_{T{\rm min}}$.
We plot separately the contributions from the two subprocesses,
$qp\rightarrow qgp$ and $gp\rightarrow ggp$.
By comparison, we also plot the cross section of 
the diffractive light quark jet.
The three curves in this figure show that the contribution from the subprocess
$gp\rightarrow ggp$ is two orders of magnitude larger than that from
the subprocess $qp\rightarrow qgp$ for the diffractive gluon jet production,
and the light quark jet production rate is in the same order with that of the
subprocess $qp\rightarrow qgp$.
This indicates that the diffractive dijet production at hadron colliders
dominantly comes from the subprocess $gp\rightarrow ggp$ in the two-gluon
exchange model.

In Fig.~8, we plot the differential cross section $d\sigma/dt|_{t=0}$ as
a function of the lower bound of the momentum fraction of the proton
carried by the incident gluon $x_{1{\rm min}}$,
where we set $k_{T{\rm min}}=5~GeV$.
Fig.5(a) is for the contribution from the subprocess
$qp\rightarrow qgp$, and Fig.5(b) is
from the subprocess $gp\rightarrow ggp$.
These two figures show that the dominant
contribution comes from the region of $x_1\sim 10^{-2}-10^{-1}$ for the
subprocess $gp\rightarrow ggp$, and $x_1>10^{-1}$ for the subprocess
$qp\rightarrow qgp$.
These properties are similar to those of the diffractive charm jet
and $W$ boson productions calculated in\cite{charm,dy}.

\section{Conclusions}

In this paper, we have calculated the diffractive light quark jet and
gluon jet productions
at hadron colliders in perturbative QCD by using the two-gluon exchange model.
We find that the production cross section is related to the squared
differential gluon distribution function $\partial G(x;Q^2)/\partial ln Q^2$
at the scale of $Q^2\sim k_T^2$, where $k_T$ is the transverse momentum of
the final state quark or gluon jet.
We have also estimated the production rates of the light quark jet and
gluon jet in the diffractive processes at the Fermilab Tevatron.

As we know, the large transverse momentum dijet production in the diffractive
processes at hadron colliders is important to study the diffractive mechanism and
the nature of the Pomeron. The CDF collaboration at the Fermilab Tevatron
have reported some results on this process\cite{tev}.
Up to now, we have calculated all of the dijet production subprocesses
in the diffractive processes at hadron colliders, including $gp\rightarrow
q\bar qp$, $qp\rightarrow qgp$ and $gp\rightarrow ggp$ processes.
In a forthcoming paper, we will compare the available data on the diffractive
dijet production cross sections at the Tevatron\cite{tev} with the predictions of
our model to test the validity of perturbative QCD description of the diffractive
processes at hadron colliders.

\acknowledgments
This work was supported in part by the National Natural Science Foundation
of China, the Education Commission Ministry of China, and the State
Commission of Science and Technology of China.


\newpage
\vskip 10mm
\centerline{\bf \large Figure Captions}
\vskip 1cm
\noindent
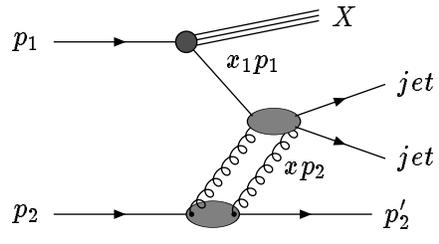
Fig.1. Sketch diagram for the diffractive dijet production at hadron colliders
in perturbative QCD. The final state jet lines represent the outgoing quark
or gluon lines, and the incident parton from the upper proton (labeled by $x_1p_1$)
can be quark or gluon correspondingly.

\noindent
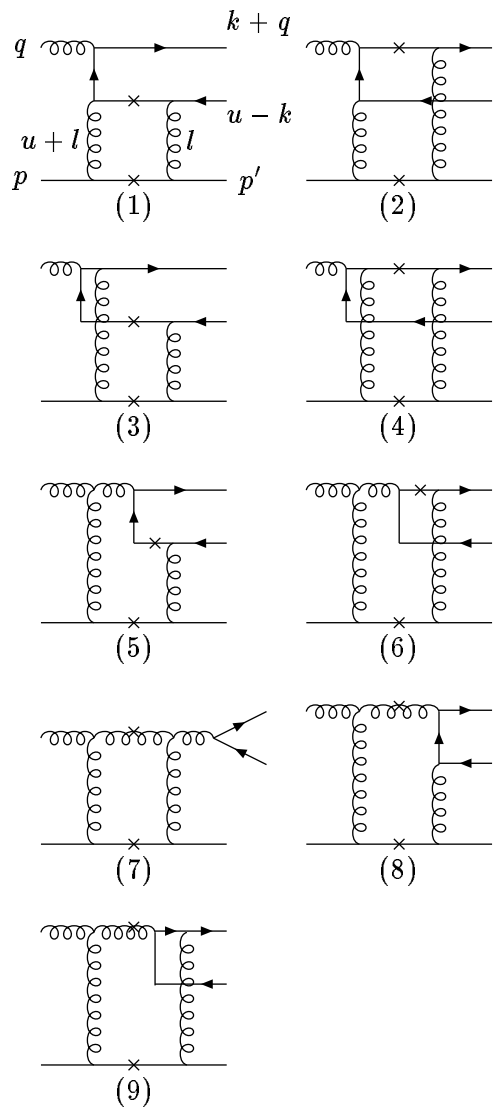
Fig.2. The lowest order perturbative QCD diagrams for partonic process
$gp\rightarrow q\bar q p$.

\noindent
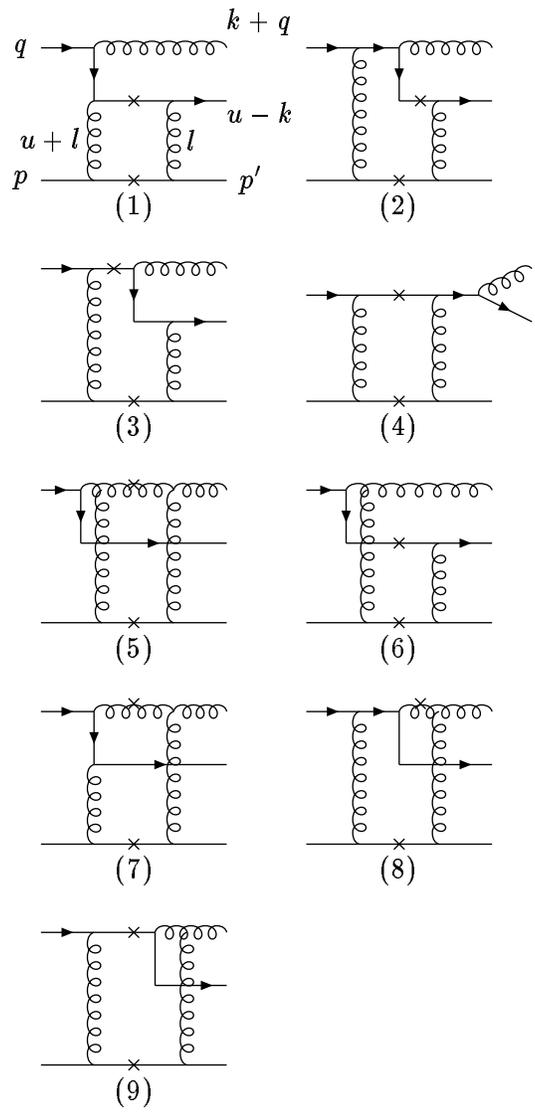
Fig.3. The lowest order perturbative QCD diagrams for partonic process
$qp\rightarrow qg p$.

\noindent
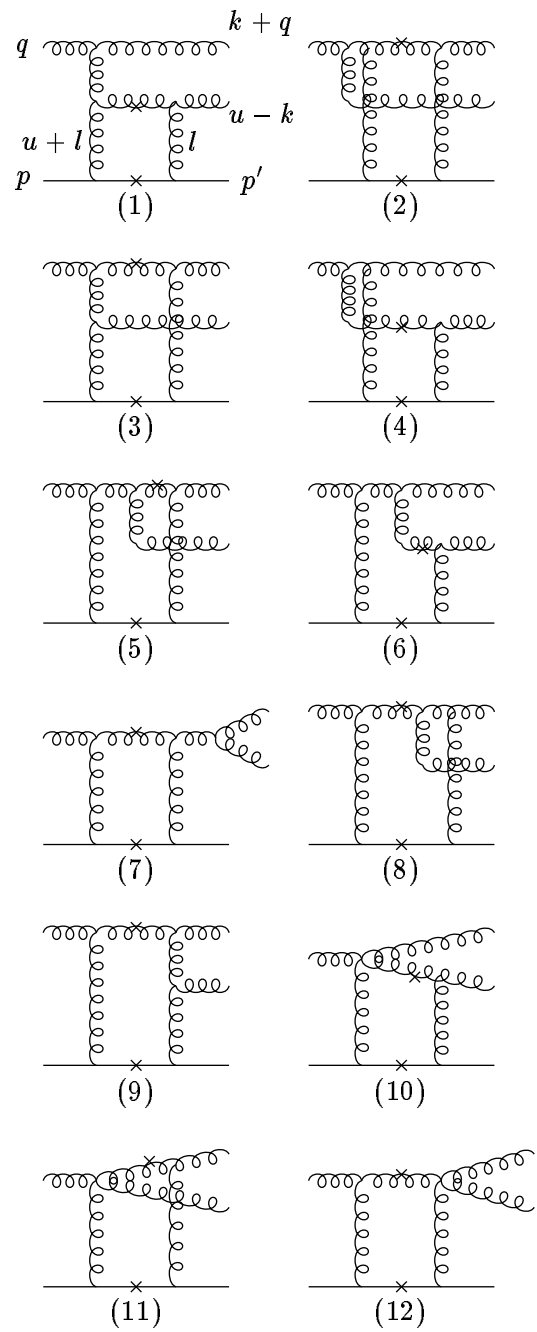
Fig.4. The lowest order perturbative QCD diagrams for partonic process
$gp\rightarrow gg p$.

\noindent
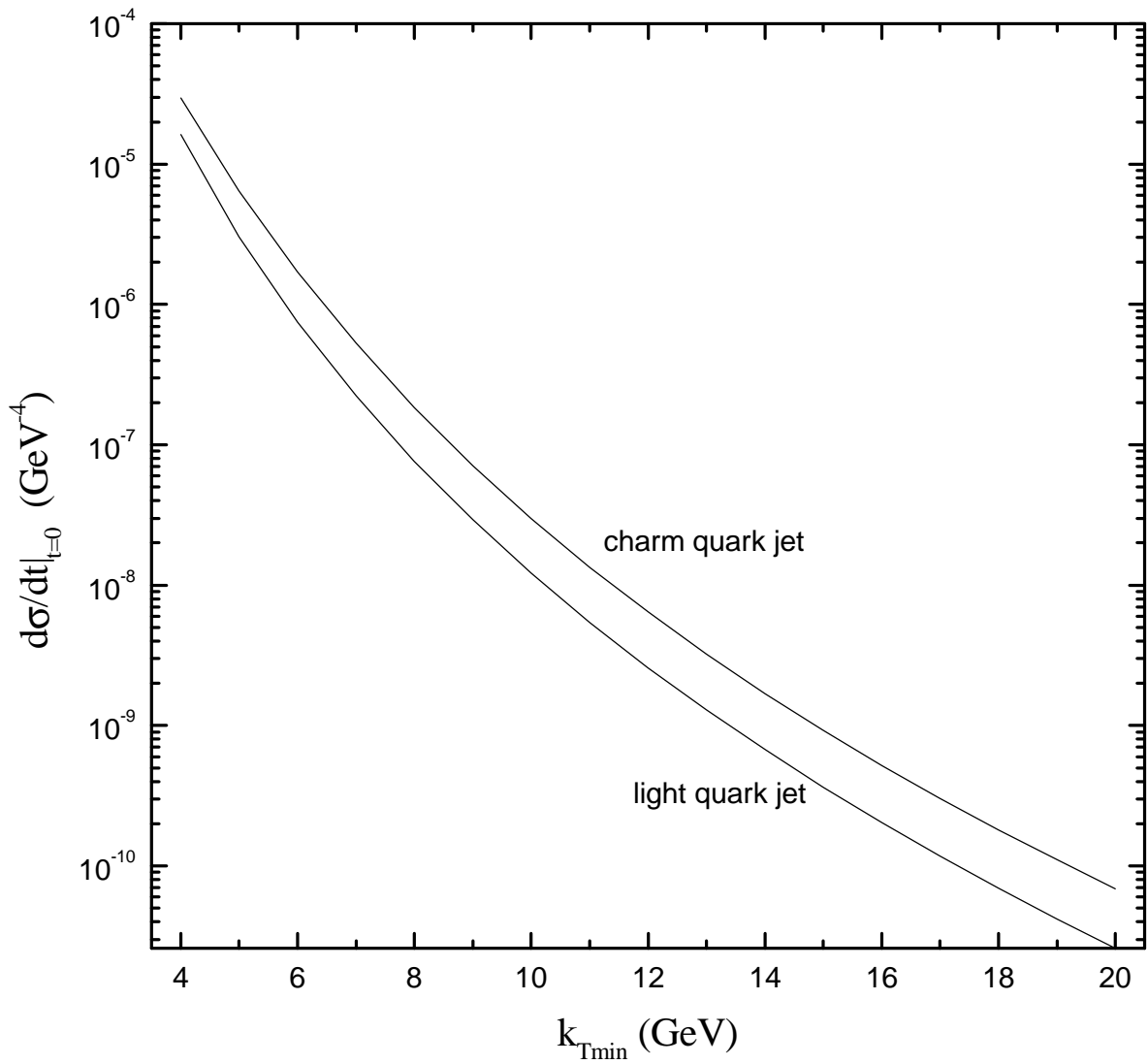
Fig.5. The differential cross section $d\sigma/dt|_{t=0}$ for the light quark jet
production in the diffractive processes as a function
of $k_{T{\rm min}}$ at the Fermilab Tevatron,
where $k_{T{\rm min}}$ is the lower bound of the transverse momentum of the
out going quark jet.
For the charm quark jet production cross section formula, we take from Ref.\cite{charm}
and set $m_c=1.5~GeV$.

\noindent
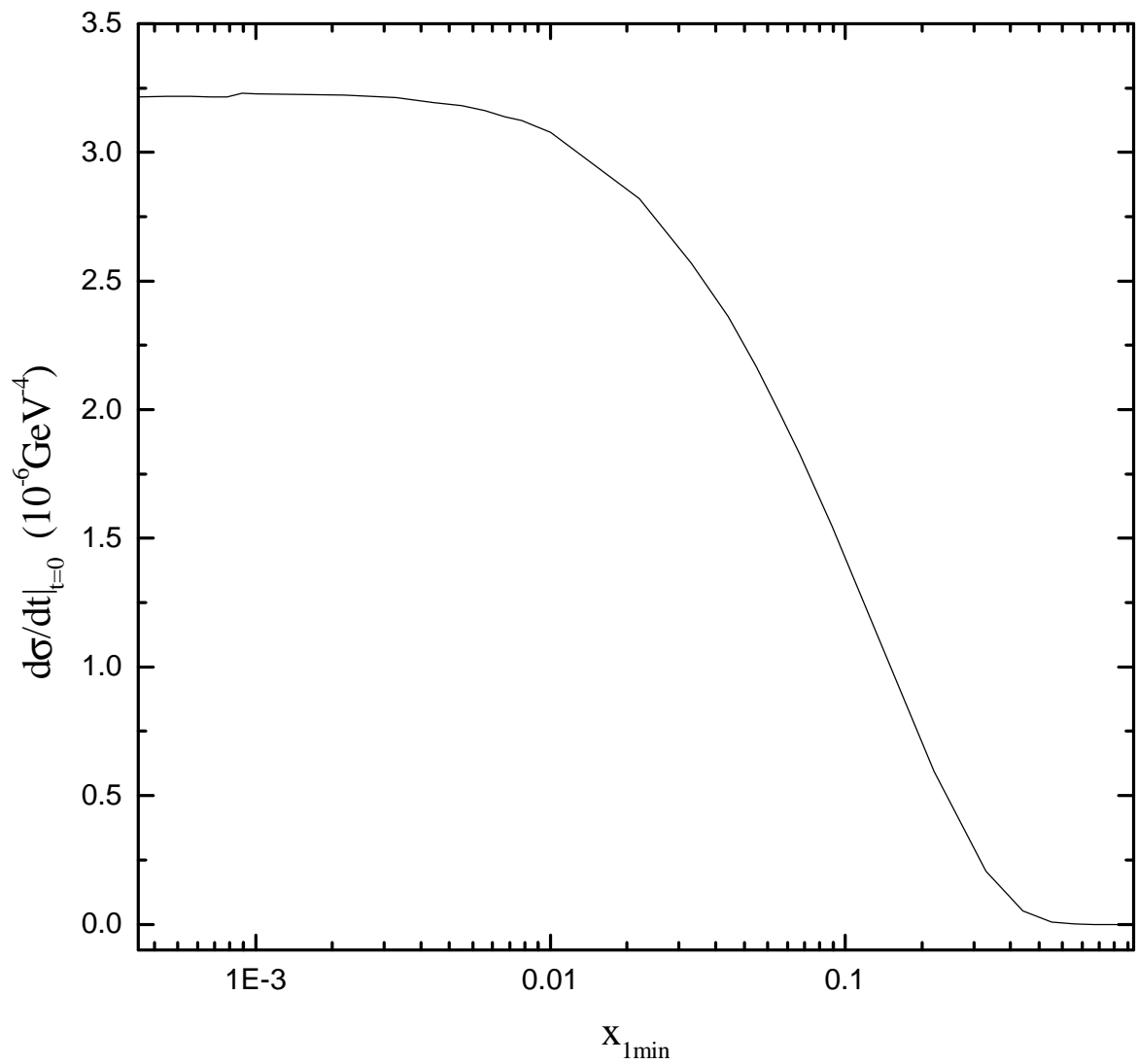
Fig.6. The differential cross section $d\sigma/dt|_{t=0}$ for the light quark jet
production 
as a function of $x_{1{\rm min}}$, where $x_{1{\rm min}}$ is the lower
bound of $x_1$ in the integration of the cross section.

\noindent
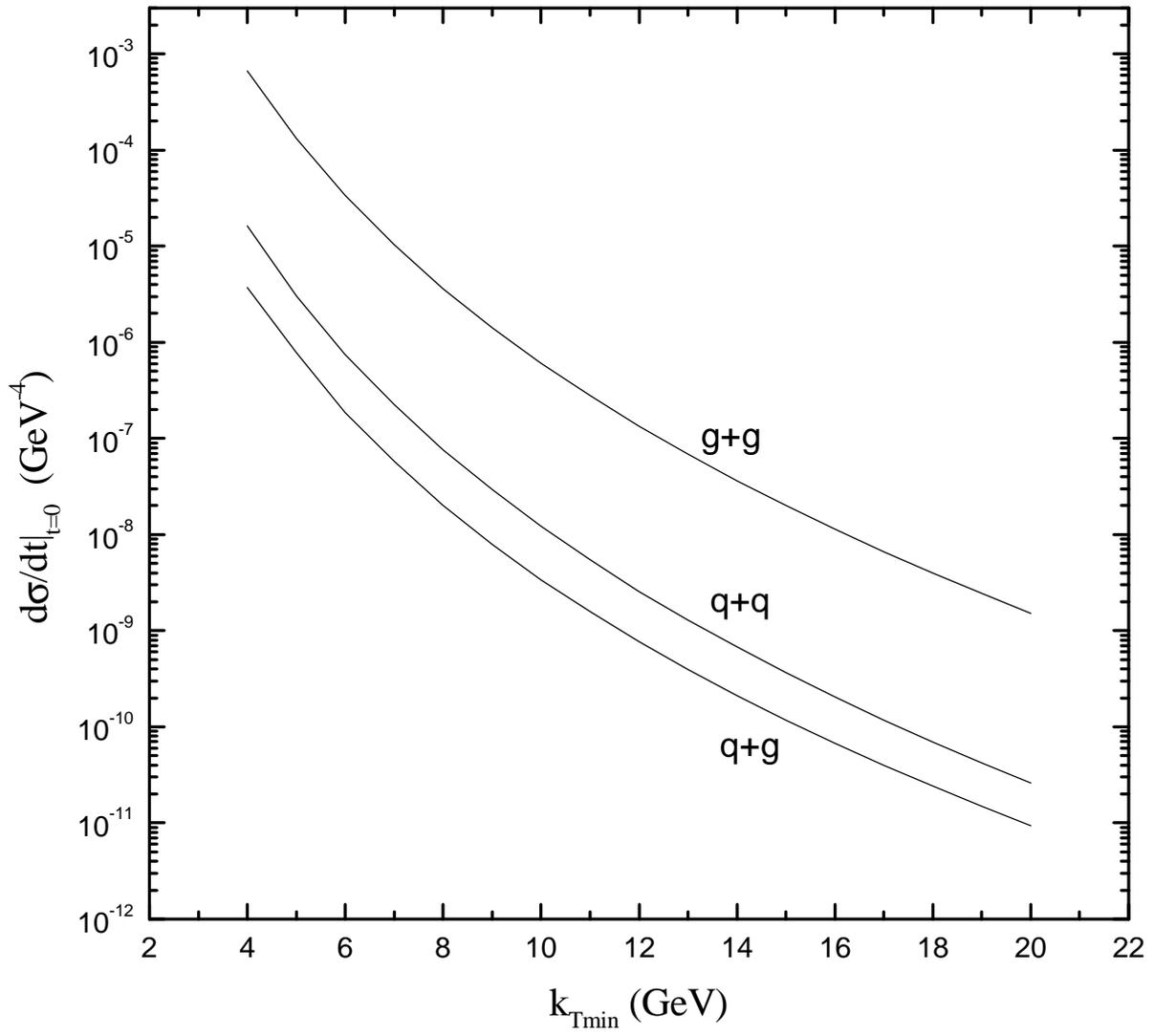
Fig.7. The differential cross section $d\sigma/dt|_{t=0}$ for the gluon jet
production in the diffractive processes as a function
of $k_{T{\rm min}}$.

\noindent
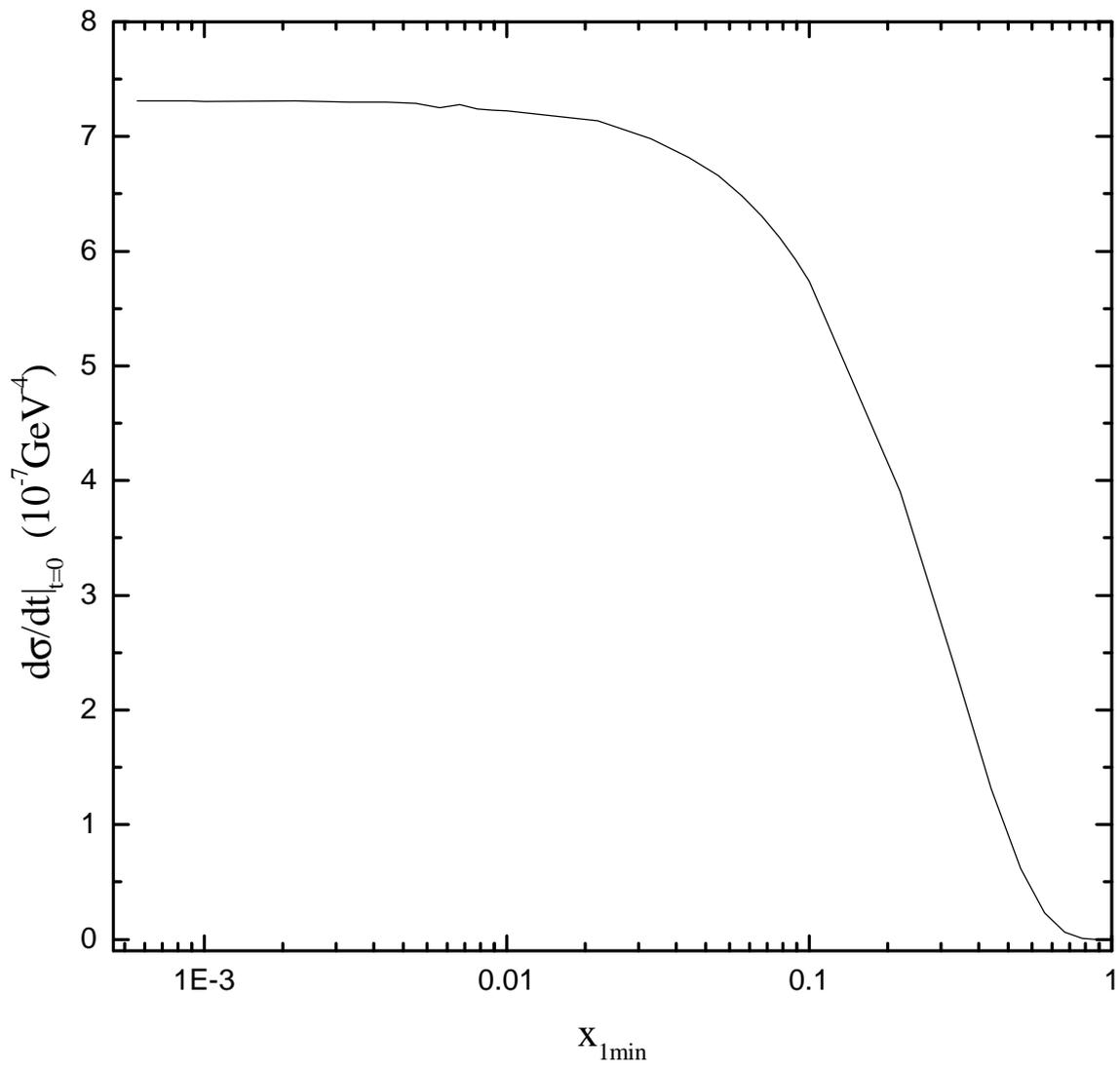
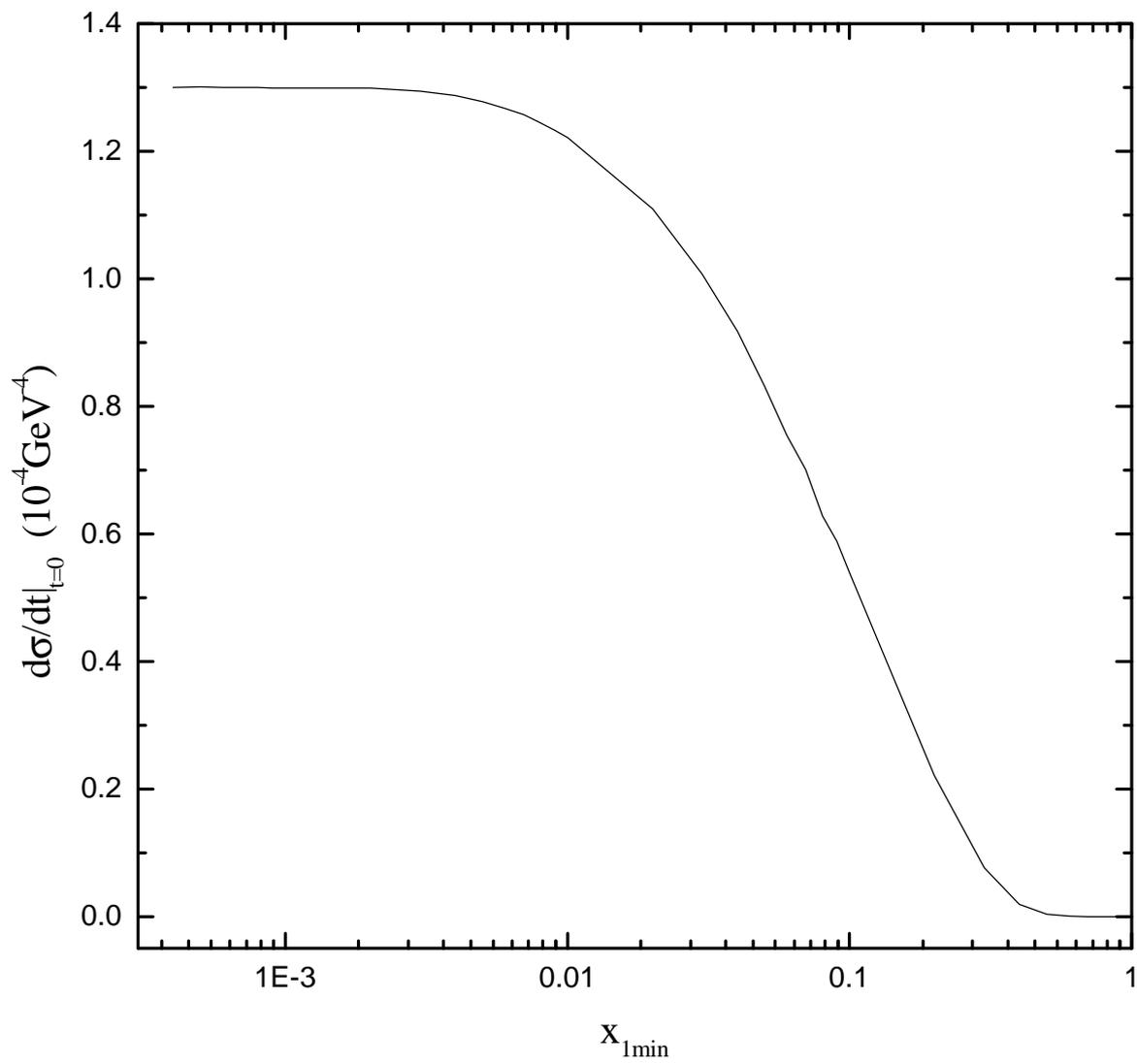
Fig.8. The differential cross section $d\sigma/dt|_{t=0}$ for the gluon jet
production
as a function of $x_{1{\rm min}}$, where $x_{1{\rm min}}$ is the lower
bound of $x_1$ in the integration of the cross section.
(a) is for the contribution from the subprocess
$qp\rightarrow qgp$, and (b) is
from the subprocess $gp\rightarrow ggp$.

\end{document}